\documentclass[%
 reprint,
 aps,               
 longbibliography,  
 amsmath,amssymb,
]{revtex4-2}
\usepackage{amsmath}
\usepackage{tipa}
\usepackage{bbm}
\usepackage{txfonts}
\usepackage{graphicx}
\usepackage{dcolumn}
\usepackage{bm}
\usepackage{amssymb}
\usepackage{latexsym}
\usepackage{color}
\usepackage{ulem}
\usepackage[mathscr]{euscript}
\usepackage[colorlinks=true,linkcolor=blue,citecolor=blue,urlcolor=blue,]{hyperref}
\usepackage[doipre={doi:~}]{uri}

\begin{document}

\title{Generalized Bloch Oscillations of Doublon Defect in One-dimensional Antiferromagnetic Atomic Spin Chain under Linear and Quadratic Magnetic Gradients}

\author{Xiang Gao$^{1}$} \email[E-mail: ]{gaoxiang@zknu.edu.cn}
\author{Xing Deng$^2$} \email[E-mail: ]{Dengxing@huat.edu.cn}
\author{Wei Xiao$^1$}
\author{Chunhua Zhang$^1$}
\author{Xingxing Wang$^1$}
\author{Xueting Fang$^{3}$}
\affiliation{$^1$School of Physics and Telecommunications Engineering, Zhoukou Normal University, Zhoukou 466001, China \\
475004,China \\
$^2$Shiyan Key Laboratory of Quantum Information and Precision Optics, and School of Mathematics, Physics and Optoelectronic Engineering, Hubei University of Automotive Technology, Shiyan 442002, China\\
$^3$MOE Key Laboratory of Fundamental Physical Quantities Measurement and Hubei Key Laboratory of Gravitation and Quantum Physics,
PGMF and School of Physics, Huazhong University of Science and Technology, Wuhan 430074, China\\
}


\begin{abstract}
We investigate generalized Bloch oscillations of a single doublon defect in a  one-dimensional antiferromagnetic $t$-$J_{\rm{z}}$ chain subjected to  linear and quadratic magnetic field gradients. 
By mapping the spin-charge coupled many-body dynamics onto an effective single-particle model, we reveal how the N\'eel background mediates parity-dependent coupling between the defect and the external fields. 
The linear gradient induces a staggered potential that splits the Wannier-Stark ladder into two intertwined subladders, producing a characteristic beating pattern in the center-of-mass motion with a subband offset accurately captured by third-order perturbation theory. 
The quadratic gradient generates sublattice-dependent tilts that shear the ladder into a scissor-like profile, yielding site-dependent beating periods that encode the  nonlocal spatial averaging of the defect wave function. 
When both gradients coexist, their interplay is fundamentally nonadditive: the quadratic gradient breaks the two-site periodicity that forbids second-order linear responses, unlocking a dominant cross-coupling proportional to $B_{\rm{z,1}}B_{\rm{z,2}}$. 
This leaves distinct fingerprints on the Bloch-oscillation spectrum---a rigid baseline  hift of the fundamental doublet and asymmetric broadening of the first harmonic. 
These predictions are quantitatively corroborated by exact many-body simulations of the real-time density evolution and center-of-mass dynamics. 
Our work establishes a gradient-induced unlocking mechanism for multi-gradient defect spectroscopy in quantum magnets, offering a framework for probing parity-protected selection rules and their controlled breaking via generalized Bloch oscillations in tunable quantum simulators.
\end{abstract}


\maketitle

\section{Introduction}
\label{section:I}
The characterization of non-equilibrium dynamics in low-dimensional quantum systems encompasses a broad class of fundamental phenomena, from anomalous transport to localization \cite{Giamarchi2003book, PolkovnikovRMP2011, AbaninRMP2019}.
Within this framework, Bloch oscillations (BOs) serve as a paradigmatic example \cite{Bloch1929, Houston1940PR, Zener1934}, describing the periodic motion of a quantum particle in a periodic potential subjected to a constant external force. 
In a tilted lattice, the continuous energy spectrum collapses into a discrete set of equally spaced Wannier-Stark (WS) ladders, leading to characteristic localized oscillations \cite{Kohn1959PR, Wannier1960PR}. 
Experimentally, BOs have been realized in semiconductor superlattices \cite{Esaki1970IBM, Feldmann1992PRB}, ultracold atomic gases \cite{Ben1996PRL, Wilkinson1996PRL},  photonic waveguide arrays \cite{Trompeter2006PRL, Lin2007apl} and acoustic waveguide arrays \cite{Moleron2019PRB, Liu2021APE}, extending the study of coherent localization to defect-free, synthetic quantum systems.

Beyond the scenario of single-particle BOs in a linear potential, extensions have been pursued along two complementary directions.
At the single-particle level, 
spatially modulated lattices with long-range hopping modify Wannier-Stark localization and induces breathing modes with symmetry-broken density profiles \cite{Efremidis2002PRE, TANG20173701PLA}, 
while temporally driven potentials generate super BOs with renormalized effective tunneling and diverging oscillation amplitudes \cite{Haller2010PRL, Kudo2011PRA}.
Crucially, higher-order gradients \cite{Zhu2019PRA} or staggered potentials \cite{Dreisow2009PRL,Maksimov2015PRA} modulate the BOs spectrum by splitting the underlying Wannier-Stark ladder into subladders with distinct level spacings, imprinting characteristic beating patterns on the real-space dynamics.
In parallel, many-body interactions induce fractional BOs of bound composite particles with suppressed effective hopping \cite{Khomeriki2010PRA, Cai2021PRL}, 
whereas quasiparticle excitations in quantum spin chains (magnons \cite{Liu2019PRA, Kosevich2013JPCM, Xu2025NC}, magnetic kinks \cite{Kyriakidis1998PRB, Hansen2022NC, Kolezhuk2024PRB}) exhibit Bloch oscillations, providing a platform for probing quantum correlations. 
Notably, BOs in spin chains have been limited to delocalized quasiparticles, leaving defect BOs under higher-order gradients and their associated beating geometries and site-dependent periodicities largely unexplored.

In this work, we investigate generalized Bloch oscillations (GBOs) of a doublon defect in a one-dimensional antiferromagnetic $t$-$J_{\rm{z}}$ atomic spin chain under linear and quadratic magnetic field gradients. 
Mapping the many-body dynamics onto an effective single-particle model reveals how the N\'eel background imprints parity-dependent structures on the defect's periodic motion: 
the linear gradient generates a characteristic beat pattern via Wannier-Stark subladder splitting with offset $\delta$; 
the quadratic gradient produces site-dependent oscillation periods by shearing the ladder into branches with slope $k_{\rm eff}=B_{z,2}\mathcal{J}_0(4t/g)$; 
and their coexistence unlocks nonadditive cross-coupling $\propto B_{z,1}B_{z,2}$ that biases the fundamental doublet and produces asymmetric broadening of the first harmonic. 
Exact many-body simulations confirm these predictions, establishing multi-gradient defect spectroscopy as a probe of parity-protected selection rules in quantum simulators.

This paper is organized as follows. In Sec.~\ref{section:II}, we introduce the model Hamiltonian and derive the effective single-particle description for the doublon defect. 
Sec.~\ref{section:III} presents our main results on GBOs under linear, quadratic, and combined field gradients. 
Sec.~\ref{section:IV} discusses the experimental feasibility of the model, and a brief conclusion is given in Sec.~\ref{section:V}; technical details are provided in the Appendices.

\section{Set up and Hamiltonian}
\label{section:II}

\begin{figure}[t]
	\centering
	\includegraphics[trim=0 20 -15 -20, width=0.5 \textwidth]{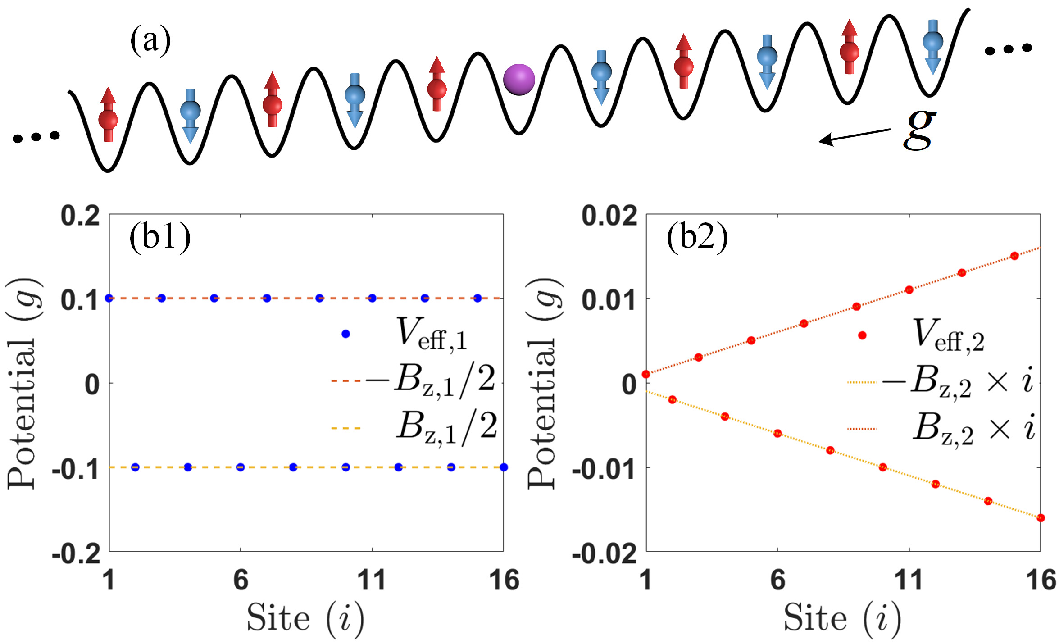}
	\caption{
    \label{fig1}
    (a) Schematic of a doublon defect (purple sphere) in Néel state. 
    (b1) The staggered potential $V_{\mathrm{eff},1}(i)=(-1)^{i+1}B_{z,1}/2$ induced by the linear gradient $B_{\rm{z,1}}$.
    (b2) The piecewise-linear potential $V_{\mathrm{eff},2}(i)=(-1)^{i+1}B_{z,2}\times i$ with sublattice-dependent slopes induced by the quadratic gradient $B_{\rm{z,2}}$. 
	Parameters: $N=16$, $g=1$, $t=0.5g$, $B_{z,1}=0.2g$, $B_{z,2}=0.01g$.}
\end{figure}

In this work, we investigate a one-dimensional atomic antiferromagnetic $t$–$J_z$ chain \cite{Montorsi2020PRA} subjected to both gravitational and magnetic fields, as realizable with magnetic atoms \cite{Lu2012PRL, Aikawa2014PRL, Baier2016Science, Chomaz2023RPP} or Rydberg-dressed atom arrays \cite{Zeiher2016NPHYS, Zeiher2017PRX}, where the Ising interaction is truncated to nearest-neighbor form by tuning the lattice spacing to the Rydberg blockade radius. 
The Hamiltonian is given by
\begin{equation}
    \begin{aligned}
    {\hat H}= &-t\sum\noindent_{\langle i,j \rangle }^{N}  \sum\noindent_{\sigma=\uparrow,\downarrow} \hat{a}_{i,\sigma}^{\dagger} \hat{a}_{j,\sigma}  
     + U\sum\noindent_{i=1}^N \hat{n}_{i,\uparrow} \hat{n}_{i,\downarrow}
    + J_{z} \sum\noindent_{ \langle i,j \rangle } ^N\hat{S}_i^z \hat{S}_j^z    \\
    &+ \sum\noindent_{i=1}^{N} \left( g \times i \right) \hat{n}_{i} + \sum\noindent_{i=1}^{N} \left( B_{\rm{z,1}} \times i + B_{\rm{z,2}} \times i^2 \right) \hat{S}_{i}^z,
    \end{aligned}
    \label{eq1}
    \end{equation}
\noindent
where $\hat{a}_{i,\sigma}^{\dagger}\left(\hat{a}_{i,\sigma}\right)$ denotes the creation (annihilation) operator of an atom with spin $\sigma$ on site $i$.
$\hat{n}_{i,\sigma}\equiv\hat{a}_{i,\sigma}^{\dagger}\hat{a}_{i,\sigma}$ and
$\hat{n}_{i}\equiv\hat{n}_{i,\uparrow}+\hat{n}_{i,\downarrow}$ count the
spin-resolved and total atom numbers at site $i$, respectively.
$\hat{S}_{i}^{z}=\frac{1}{2}\left(\hat{n}_{i,\uparrow}-\hat{n}_{i,\downarrow}\right)$
is the spin operator and $\langle i,j\rangle$ indicates the summation over the
nearest neighbors in the lattice.
The first three terms in $\hat{H}$ describe the nearest-neighbor (NN) hopping,
the repulsive on-site Hubbard interaction, and the NN Ising interaction between
spins, where $t$, $U>0$, and $J_z>0$ are the spin-independent tunneling
strength, on-site interaction strength, and antiferromagnetic interaction
strength, respectively.
The last two terms in $\hat{H}$ account for external potentials due to
gravitational and magnetic fields, where $g$, $B_{\rm{z,1}}$, and $B_{\rm{z,2}}$
denote the strengths of the linear gravitational potential and the linear and
quadratic magnetic field gradients, respectively.
Unless otherwise stated, we fix $U=200g$ and $J_z=100g$ throughout this work,
i.e., $U=200$ and $J_z=100$ for $g=1$.
The on-site interaction contributes only a constant energy offset within the
truncated single-doublon subspace and therefore does not affect the projected
defect dynamics; it becomes relevant for virtual states outside this subspace,
where it determines the additional energy cost of extra hole--doublon pairs.
We focus on the parameter regime $J_z\gg t$
\cite{Qiao2025Nature, Zhang2026PRL}, and the hierarchy
$J_z\gg B_{\rm{z,1}}\gg B_{\rm{z,2}}$, ensuring that the magnetic gradients act
as weak perturbations on the antiferromagnetic background while the doublon
exhibits coherent dynamics.

At half filling and in the absence of external fields, the strong Ising interaction favors an antiferromagnetic N\'eel state, which serves as the classical background for the defect dynamics considered below.
The gravitational field couples only to the charge (density) sector and the magnetic fields considered here are too weak to excite spinons.
Building on this, we insert a defect, such as a doublon (a doubly occupied site) , into the Néel state $ \left| \uparrow  \downarrow \cdots \uparrow \downarrow \right\rangle $ as illustrated in Fig.~\ref{fig1}(a).
The many-body dynamics can then be restricted to a truncated, energetically isolated Hilbert subspace $\mathcal{H}=\{ \left| i \right\rangle   \}$, in which the doublon moves while the spin background remains in its classical Néel configuration. 
The basis states are defined as 
\begin{equation}
    \begin{aligned}
    \left| i \right\rangle =  \begin{cases}
    \left| \uparrow_1  \downarrow_2 \cdots  \downarrow_{i-1} D_i \uparrow_{i+1} \downarrow_{i+2}   \cdots  \downarrow_{N}  \right\rangle   \quad \quad \ i \ \  {\rm{  \ odd}} \\
    \left| \uparrow_1  \downarrow_2 \cdots \downarrow_{i-2}   \uparrow_{i-1} D_i \downarrow_{i+1} \cdots  \downarrow_{N}  \right\rangle    \quad \quad \ i \ \ {\rm{  \ even}}   
    \end{cases},
    \end{aligned}
    \label{eq2}
    \end{equation}
\noindent
where $D_i$ marks a doublon at site $i$. 
Within $\mathcal{H}$, the original spin-charge coupled many-body dynamics projects onto a leading-order effective single-particle model: a quasiparticle with hopping amplitude $t$ subject to position-dependent potential $V_{\rm{eff}}$ that encodes both the external field gradients and underlying spin order. The effective Hamiltonian reads
\begin{equation}
    \begin{aligned}
    {\hat H}_{\rm{eff}}= &-t \sum\noindent_i^N \left( \left| i \right\rangle \left\langle i+1 \right| +  \left| i+1 \right\rangle \left\langle i \right|   \right)  \\
    &+ \sum\noindent_{i=1}^N  \left( g \times i +  V_{\rm{eff},1} \left( i \right) + V_{\rm{eff},2} \left( i \right)  \right) \left| i \right\rangle \left\langle i \right|   + E_0,
    \end{aligned}
    \label{eq3}
    \end{equation}
\noindent
where $V_{\rm{eff,1}} \left( i \right)$ and $V_{\rm{eff,2}} \left( i \right)$ are site-dependent potentials derived from the coupling between the external fields and the underlying Néel order. Explicitly,
\begin{equation}
    \begin{aligned}
    V_{\rm{eff,1}} \left( i \right) =  \frac{(-1)^{i+1} B_{\rm{z,1}}}{2}
    \end{aligned}
    \label{eq4}
    \end{equation}
\noindent
and
\begin{equation}
    \begin{aligned}
     V_{\rm{eff,2}} \left( i \right) = (-1)^{i+1} B_{\rm{z,2}} \times i,
    \end{aligned}
    \label{eq5}
    \end{equation}
\noindent
The constant $E_0=  U +     g N\left( N+1 \right) /2 - B_{\rm{z,1}} \left( N -1 \right)/2  - B_{\rm{z,2}} N \left( N + 1 \right)/2 $ 
collects  the uniform energy shift from the external potentials. It depends on the chain length $N$ and the field strengths, but does not affect the defect’s relative motion or spectral properties. 

As illustrated in Fig.~\ref{fig1}(b1) and (b2), the spatial profiles of $V_{\rm{eff,1}} \left( i \right)$ and $V_{\rm{eff,2}} \left( i \right)$ exhibit a piecewise-linear structure that depends explicitly on the site parity. 
This parity dependence is a direct consequence of the staggered spin polarization in the Néel background, which modulates the effective Zeeman shift induced by the magnetic gradients.
In particular, the linear magnetic gradient $B_{\text{z,1}}$ manifests as a constant energy offset between adjacent sites in $V_{\text{eff,1}}(i)$, while the quadratic magnetic gradient $B_{\text{z,2}}$ yields both sublattice-dependent slopes. 
The competition between the kinetic hopping $t$ and these tailored potential landscapes governs the defect's motion, leading to GBOs whose real-space periodicity is modulated by spectral splitting and shearing of the underlying Wannier-Stark ladder.

The effective potentials $V_{\rm eff,1}(i)$ and $V_{\rm eff,2}(i)$ entering Eq.~\eqref{eq3} are weakly renormalized by virtual hopping processes that transiently take the system outside the truncated Hilbert space $\mathcal{H}$.
As shown in Appendix~\ref{app-E}, these processes create intermediate hole--doublon or doublon--hole pairs and generate a second-order correction to the effective on-site potential,
\begin{equation}
    \delta V_{\rm leak}(i)
    =
    (-1)^{i+1}\kappa_{\rm eff}B_{z,2}
    +
    B_{\rm cross}\,i\,B_{z,1}B_{z,2},
    \label{eq:leak-main}
\end{equation}
where $\kappa_{\rm eff}\sim t^2/\Delta_{\rm leak}^2$ and $B_{\rm cross}\sim t^2/\Delta_{\rm leak}^3$, with $\Delta_{\rm leak}=U+J_z/2$; all site-independent contributions are absorbed into $E_0$.
Equivalently, to this order the leakage correction can be absorbed into the parameter renormalizations
$g\rightarrow g+B_{\rm cross}B_{z,1}B_{z,2}$ and
$B_{z,1}\rightarrow B_{z,1}+2\kappa_{\rm eff}B_{z,2}$,
while $B_{z,2}$ remains unchanged.
For the parameters used throughout this work, $U=200g$, $J_z=100g$ and $t=0.5g$, the characteristic leakage gap is $\Delta_{\rm leak}=250g$ and hence $t/\Delta_{\rm leak}=1/500$.
The virtual leakage is therefore controlled by the small parameter $(t/\Delta_{\rm leak})^2\simeq4\times10^{-6}$, leaving the structure of Eqs.~\eqref{eq3}--\eqref{eq5} unchanged and producing only weak renormalizations of the effective parameters.
This is directly confirmed by full-space evolution, which yields a maximum leakage weight $\mathcal{L}_{\max}=8.7\times10^{-5}$ at these parameters [Fig.~\ref{figE2}(a)]. 
We accordingly adopt Eqs.~\eqref{eq3}--\eqref{eq5} as the leading-order effective description of the doublon dynamics.
This projection remains controlled provided that $t$ and the characteristic field scales $g$, $B_{z,1}$, and $B_{z,2}$ are small compared with $\Delta_{\rm leak}$.
The projection is expected to break down when any of these scales becomes comparable to $\Delta_{\rm leak}$, such that real hole--doublon and spinon excitations are no longer energetically suppressed.

In the following, we investigate these phenomena using a combined numerical and analytical approach. 
The numerical simulations are based on the many‑body Hamiltonian $\hat{H}$ (Eq.~\ref{eq1}) for a finite chain of length $N=16$ with open boundaries, initialised in the classical N\'eel state containing a single doublon at the central site, i.e., $\left| \psi_{\rm{ini}} \right\rangle= \left| i=8 \right\rangle$. 
Note that $N=16$ with central injection is effectively free of finite-size effects: the Wannier--Stark-localized doublon explores only $\sim4t/g$ sites around the injection point and never reaches the boundaries within the evolution window, and the leakage analysis of Appendix~\ref{app-E} holds for arbitrary $N$ with boundary bonds accounted for separately [Fig.~\ref{figE2}(a) uses the same $N=16$ chain].
We track the site-resolved density evolution $\rho_i  \left( t \right)=\left\langle \psi \left( t \right) \left| \hat{n}_i  \right|  \psi \left( t \right)  \right\rangle$ and the center-of-mass trajectory $ \left\langle x_{\rm{c.m.}} \left( t \right) \right\rangle = \left\langle \psi \left( t \right) \left| \sum_i i \hat{n}_i \right|  \psi \left( t \right)  \right\rangle$ to capture the defect's periodic motion.
In parallel, we complement these simulations with the effective single-particle description based on Eq.~\ref{eq3} to interpret the numerical results analytically.
The spacing of the Wannier-Stark ladder directly encodes the Bloch-oscillation frequency: a ladder spacing $\Delta E$ sets the Bloch-oscillation period $T_B=2\pi/\Delta E$ for both $\rho_i(t)$ and $\langle x_{\rm{c.m.}}(t)\rangle$.
By resolving the generalized ladder structure that emerges under linear and quadratic gradients, we thus obtain the effective oscillation frequencies, renormalized slopes, and cross-coupling shifts of the defect's real-space Bloch oscillations, which we simultaneously exhibit in the time domain.

\section{Generalized Bloch oscillation of defect in antiferromagnetic chain}
\label{section:III}
\subsection{Parity-Induced Spectral Splitting under a Linear Magnetic Gradient}
\label{section:III-A}

\begin{figure}[t]
	\centering
	\includegraphics[trim=0 30 0 -20, width=0.48 \textwidth]{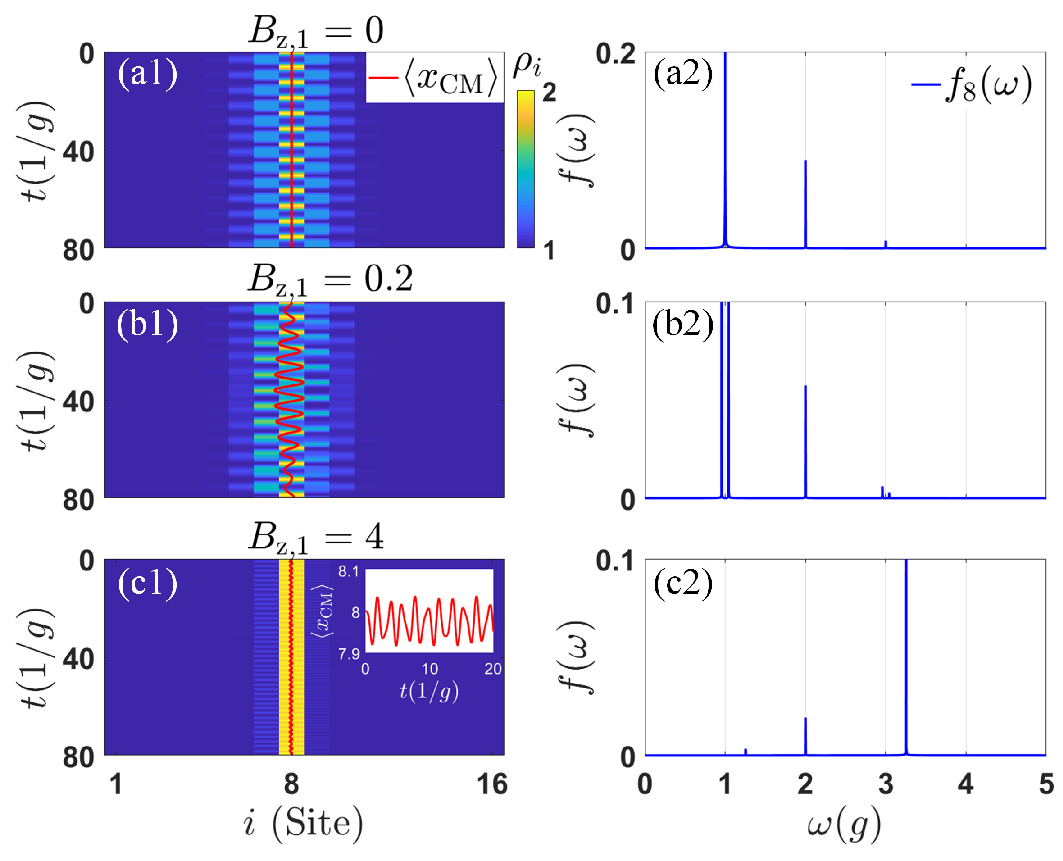}
	\caption{
    \label{fig2}
    Doublon dynamics under a linear magnetic gradient $B_{z,1}$. 
    (a1)--(c1) Time evolution of the site-resolved doublon density $\rho_i(t)$ (background colormap) and the center-of-mass trajectory $\langle x_{\mathrm{c.m.}}(t)\rangle$ (red solid line) for the initial state $|\psi_{\mathrm{ini}}\rangle=|i=8\rangle$ with $B_{z,1}=0$, $0.2g$, and $4g$ (top to bottom).
    (a2)--(c2) Corresponding Fourier spectrum $f_8(\omega)$ of the density oscillations at the initial site. 
    Parameters: $N=16$, $g=1$, $t=0.5g$, $B_{\rm{z,2}}=0$.
	}
\end{figure}

We first investigate the doublon dynamics under a linear magnetic gradient, 
tracking the real-time density evolution and center-of-mass trajectories 
across zero, weak, and strong fields. 
In the absence of a magnetic gradient ($B_{\mathrm{z,1}}=0$), the doublon 
undergoes conventional BOs, a hallmark of the Wannier-Stark localization 
in a tilted lattice. 
As shown in Fig.~\ref{fig2}(a1), the density profile oscillates symmetrically 
around the injection site $i=8$ with period $T_B=2\pi/g$ set by the inverse 
gravitational tilt, while the center of mass remains stationary throughout 
the evolution (red solid line). 
The Fourier spectrum $f_{8}(\omega)$ of the density oscillations at the initial site, shown in Fig.~\ref{fig2}(a2), exhibits equidistant peaks at integer multiples of $g$, reflecting the uniform ladder spacing that underpins the BOs.

The linear magnetic gradient modifies the doublon's BO by imposing a two-site staggered potential. 
For a weak magnetic gradient ($B_{\mathrm{z,1}} < g$), the staggered potential breaks the spatial symmetry of the density oscillations, as evidenced by the temporal evolution of the density in Fig.~\ref{fig2}(b1). 
Consequently, the center of mass $\langle x_{\rm c.m.}(t)\rangle$ exhibits a characteristic beating pattern (red line in Fig.~\ref{fig2}(b1)), reflecting the coherent superposition of two dominant BO frequencies $\omega_\pm \approx g \pm \delta$. The 
Fourier spectrum $f_{8}(\omega)$ in Fig.~\ref{fig2}(b2) reveals this frequency-domain signature: the spectral weight at odd multiples of $g$ splits into doublets, whereas even multiples remain unsplit. 
This parity-dependent spectral splitting signifies the formation of two intertwined Wannier-Stark subladders with nearest-neighbor gaps $\Delta E_1^\pm = |g \pm \delta|$.

In the strong-field regime ($B_{\mathrm{z,1}} \gg g$), the magnetic gradient dominates and suppresses the BO amplitude. 
The doublon defect executes strongly localized oscillations confined to a narrow region around the initial site (Fig.~\ref{fig2}(c1)), as the large sublattice offset freezes inter-dimer tunneling and fragments the periodic orbit into small-amplitude wiggles. 
The center of mass exhibits tiny, high-frequency oscillations (see inset). Fig.~\ref{fig2}(c2) shows that the fundamental BO frequency shifts with $B_{\mathrm{z,1}}$, tracking the frozen sublattice offset, while the next-nearest-neighbor peak remains pinned at $\omega=2g$, indicating that second-neighbor Bloch oscillations driven by the uniform gravitational tilt persist even when nearest-neighbor tunneling is frozen out.

\begin{figure}[t]
	\centering
	\includegraphics[trim=10 25 -20 0, width=0.5 \textwidth]{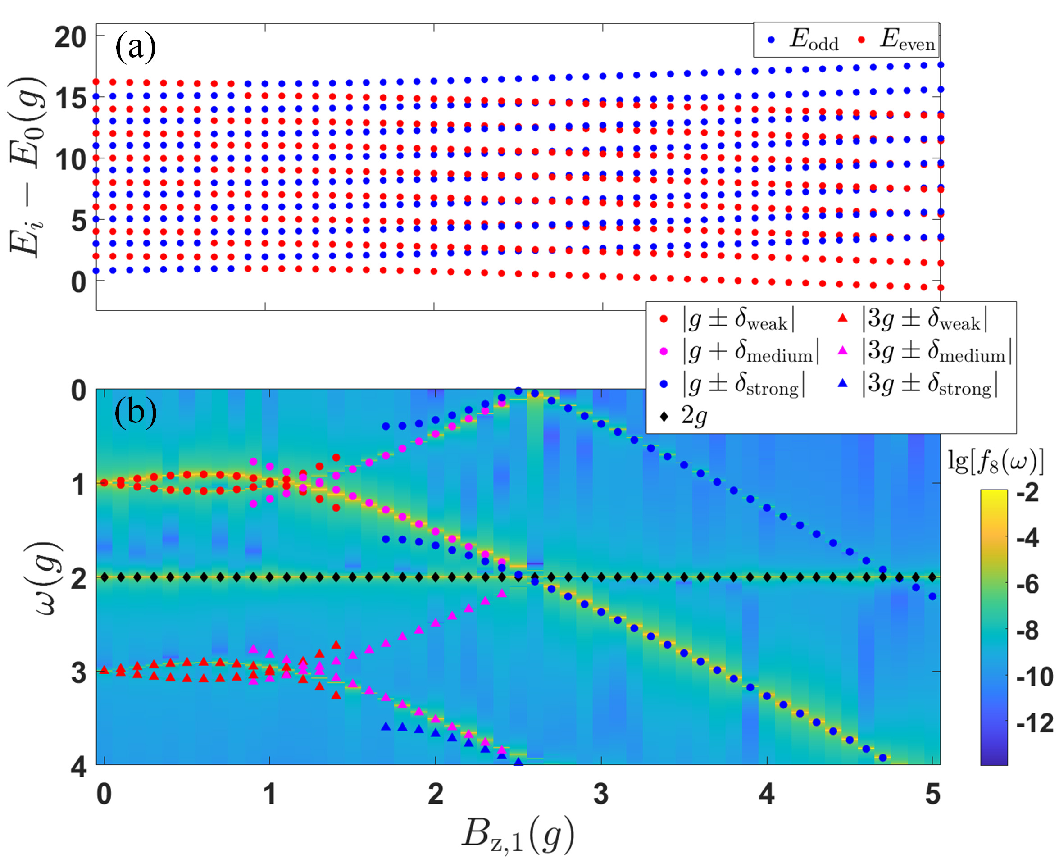}
	\caption{
    \label{fig3}
    (a) Reconstructed Wannier-Stark ladder as a function of the linear gradient $B_{z,1}$.
    Blue (red) dots denote the eigenenergies of the odd (even) subladder branches, respectively.
    (b) Fourier spectrum $f_8(\omega)$ of the doublon density oscillations at the initial site as a function of $B_{z,1}$.
    The color of the symbols indicates the theoretical regime: red for the weak-field theory ($\delta=\delta_{\mathrm{weak}}$), magenta for the medium-field theory ($\delta=\delta_{\mathrm{medium}}$), and blue for the strong-field theory ($\delta=\delta_{\mathrm{strong}}$). 
    The shape encodes the spectral peak identity: solid circles for $\omega=|g\pm\delta|$, solid triangles for $\omega=|3g\pm\delta|$, and black diamonds for the pinned next-nearest-neighbor gap $\omega=2g$.
    Parameters: $N=16$, $g=1$, $t=0.5g$, $B_{z,2}=0$.
	}
\end{figure}

The beating pattern and parity-dependent spectral splitting shown in Fig.~\ref{fig2} arises from the dimerization of the Wannier-Stark ladder into two intertwined subladders.
The staggered potential $V_{\text{eff,1}}(i)$ induces an alternating offset $\pm B_{\rm{z,1}}/2$ between adjacent sites, doubling the unit cell and folding the Brillouin zone.
This transforms the system into a binary superlattice with uniform hopping.
Under the global linear tilt $g$, this superlattice hosts two Bloch subbands \cite{Dreisow2009PRL,Maksimov2015PRA},  which manifest as two intertwined Wannier-Stark subladders with the same tilt induced by $g$ and a relative energy offset $\delta$ governed by $B_{\mathrm{z,1}}$. 
In Fig.~\ref{fig3}(a), we show the field-dependent Wannier-Stark ladder reconstruction, where the eigenenergies of the odd (blue dots) and even (red dots) subladder branches shear rigidly in opposite directions with increasing $B_{\mathrm{z,1}}$.
As shown by Maksimov et al. \cite{Maksimov2015PRA}, the eigenenergies of these subladders can be approximated as:
\begin{equation}
    \begin{aligned}
        E_{2n} &\approx 2n \cdot g - \frac{\delta}{2} + E_0, \\
        E_{2n+1} &\approx (2n+1) \cdot g + \frac{\delta}{2} + E_0,
        \end{aligned}
    \label{eq6}
    \end{equation}
\noindent
where $\delta$ is the renormalized subband offset.

Therefore, the energy gaps between adjacent eigenstates are
\begin{equation}
    \begin{aligned}
        &\Delta E_{1}^{+} =E_{2n+1}-E_{2n} =g+\delta, \\
        &\Delta E_{1}^{-} =E_{2n+2}-E_{2n+1} =g-\delta.
    \end{aligned}
    \label{eq7}
\end{equation}
These parity-dependent gaps directly dictate the GBOs spectrum shown in Fig.~\ref{fig2}.
The nearest-neighbor gaps $\Delta E_1^\pm = |g \pm \delta|$ generate the two dominant frequency components $\omega_\pm \approx g \pm \delta$ whose interference produces the beat pattern, while the next-nearest-neighbor gap $\Delta E_2 = 2g$ yields the pinned second-harmonic frequency. 
This pattern extends to all higher orders: odd-multiple peaks split into doublets at $\omega_{2n+1}^\pm = |(2n+1)g \pm \delta|$, whereas even-multiple peaks remain unsplit at $\omega_{2n} = 2ng$, directly accounting for the spectral structure in Fig.~\ref{fig2}(b2).

The dependence of the subband offset $\delta$ on $B_{\mathrm{z,1}}$ can be derived within perturbation theory.
In the weak-field regime ($B_{\mathrm{z,1}} < g$), the staggered potential $V_{\rm{eff,1}}$ is treated as a perturbation to the uniform Wannier-Stark ladder, and a third-order expansion yields:
\begin{equation}
     \delta_{\rm{weak}} \approx \mathcal{A}_1 B_{\mathrm{z,1}} + \mathcal{A}_3  B_{\mathrm{z,1}} ^3 + \mathcal{O}\left( B_{\mathrm{z,1}}^4 \right), 
    \label{eq8}
\end{equation}
\noindent
where the coefficients $\mathcal{A}_1$ and $\mathcal{A}_3$ originate from the first- and third-order corrections to the eigenenergies, respectively, with explicit forms given in Appendix \ref{app-A}.
Notably, the second-order correction vanishes exactly because the staggered potential $V_{\text{eff,1}}(i)$ is strictly two-site periodic while the unperturbed Wannier-Stark states are translationally invariant; virtual transitions to the left and right contribute with equal magnitude but opposite sign, forcing an exact cancellation. 
This symmetry-protected suppression pushes the leading nonlinear response to third order and proves essential for the non-additive physics in Sec.~\ref{section:III-C}.

As $B_{\mathrm{z},1}$ increases beyond $g$, the weak-field perturbation breaks down and the system enters a crossover regime where the local potential dominates over the kinetic energy. 
Exploiting the two-site periodicity of the staggered potential $V_{\rm{eff,1}} \left( i \right)$, we partition the chain into a series of dimers, comprising adjacent odd ($2n-1$) and even ($2n$) sites, and treat the inter-dimer tunneling as a perturbation.  
Within each dimer, the intracell dynamics is governed by the competition between the  intracell detuning ($ \sim B_{\mathrm{z},1}-g $) and the tunneling ($t$), yielding  eigenstates $|n, \pm \rangle$ with zeroth-order energies $E_{n,\pm}^{(0)} = (2n-1/2)g \pm \Delta$, where $\Delta = \sqrt{[(B_{\mathrm{z},1}-g)/2]^2 + t^2}$.
Including second-order corrections from virtual transitions between adjacent dimers, the subband offset becomes:
\begin{equation}
    \delta_{\mathrm{medium}} = 2\Delta + g - \frac{t^{2}\left[(B_{\mathrm{z},1}-g)(B_{\mathrm{z},1}-3g)+2t^{2}\right]}{2\Delta\left[(3g-B_{\mathrm{z},1})(g+B_{\mathrm{z},1})-4t^{2}\right]}.
    \label{eq9}
\end{equation}
In the strong-field limit ($B_{\mathrm{z},1} \gg g$), where $\Delta \approx (B_{\mathrm{z},1}-g)/2$ and the high-energy states become predominantly localized on odd sites, Eq.~\eqref{eq9} reduces to the compact form:
\begin{equation}
    \delta_{\mathrm{strong}} = B_{\mathrm{z},1} + \frac{2t^{2}}{g+B_{\mathrm{z},1}}.
    \label{eq10}
\end{equation}
reflecting the dominance of the bare staggered potential with weak renormalization from virtual intercell transitions.

We quantitatively validate this crossover by a systematic scan of the Fourier spectrum 
$f_8(\omega)$ versus $B_{z,1}$ up to $B_{z,1}=5g$, corresponding to $B_{z,1}/J_z=0.05$ [Fig.~\ref{fig3}(b)]. 
In the weak-field regime 
($B_{z,1}<g$), the fundamental peak splits into doublets at $\omega\approx g\pm\delta_{\rm weak}$, with the splitting initially growing linearly and then bending due to the cubic correction $\propto B_{z,1}^3/g^2$. 
As $B_{z,1}$ increases toward $g$, the lower branch $\omega=\Delta E_1^-$ softens non-monotonically, reaching a minimum where $\delta\approx g$ before rising again under $\delta_{\rm strong}\approx B_{z,1}$. 
Throughout this entire crossover, the next-nearest-neighbor gap remains pinned at $\Delta E_2=2g$ [horizontal dashed line in Fig.~\ref{fig3}(b)], because transitions within the same subladder (even$\to$even or odd$\to$odd) experience opposite sublattice shifts ($\pm\delta$) that cancel exactly, leaving only the uniform gravitational tilt $2g$.
For a purely linear gradient, leakage produces no site-dependent correction and hence no peak shift (Appendices~\ref{app-E} and~\ref{app-F}).

\subsection{Parity-induced Spectral Tilting under a quadratic magnetic field}
\label{section:III-B}
\begin{figure}[t]
	\centering
	\includegraphics[trim=0 20 -20 0, width=0.5 \textwidth]{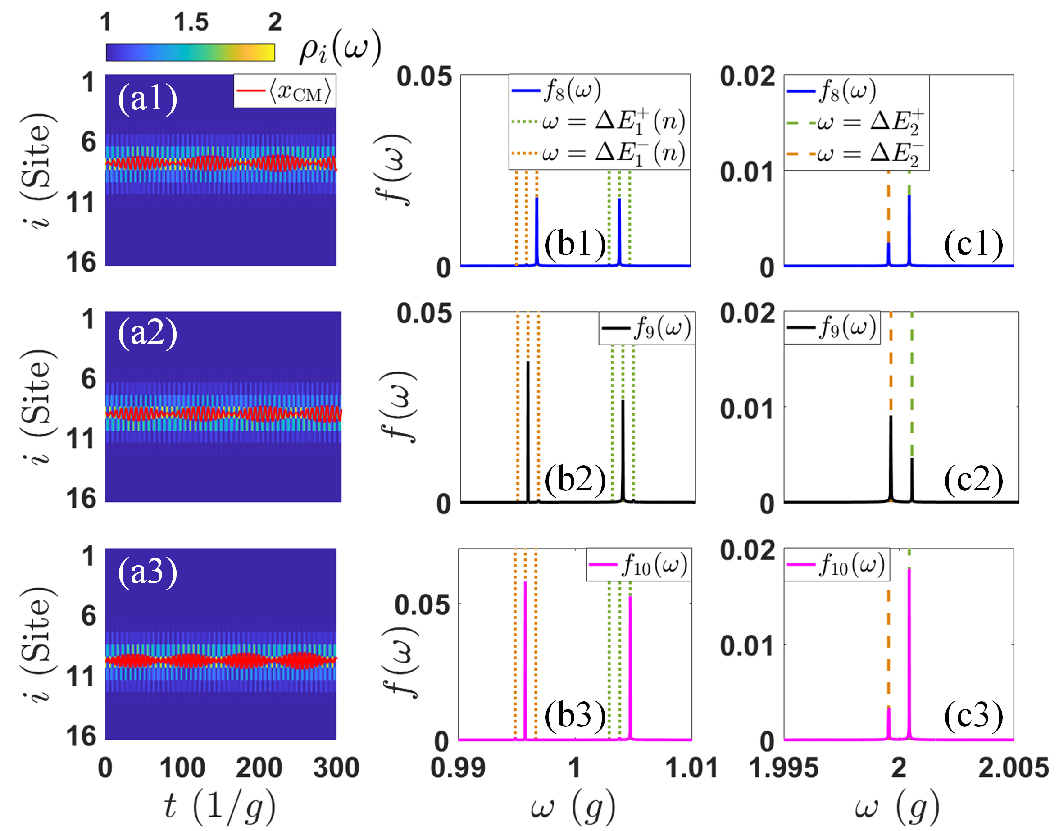}
	\caption{
    \label{fig4}
    Doublon dynamics under a purely quadratic magnetic gradient $B_{z,2}$ for different initial injection sites.
    (a1)--(a3) Time evolution of the site-resolved density $\rho_i(t)$ (background colormap) and center-of-mass trajectory $\langle x_{\mathrm{c.m.}}(t)\rangle$ (red solid line) for $i=7,8,9$ (top to bottom).
    (b1)--(b3) and (c1)--(c3) Fourier spectra $f_i(\omega)$ in the fundamental region $\omega\approx g$ and first harmonic $\omega\approx 2g$ (solid lines).
    The dotted lines and dashed lines in (b) and (c) indicate theoretical predictions from Eq.~\eqref{eq13} and Eq.~\eqref{eq14}. 
    Parameters: $N=16$, $g=1$, $t=0.5g$, $B_{z,1}=0$, $B_{z,2}=0.001g$.
	}
\end{figure}

Unlike the linear gradient, which produces a site-independent beating pattern, the quadratic gradient generates a site dependence of the oscillation periods. 
Fig.~\ref{fig4} (a1)-(a3) shows the density evolution and center-of-mass trajectories for initial sites $i=7,8,9$. 
Initialization at lower $i$ yields slow envelope modulations of the center of mass, while higher $i$ produces progressively rapid oscillations, indicating that the oscillation period shortens monotonically with the injection site. 

The corresponding Fourier spectra reveal the spectral fingerprints of this site-dependent dynamics. 
In the fundamental region $\omega\approx g$, the peaks form a bipartite comb that fans out symmetrically around $\omega=g$, with the splitting growing linearly with the initial site, as shown in Fig.~\ref{fig4}(b). 
In contrast, the first-harmonic region $\omega\approx 2g$ [Fig.~\ref{fig4}(c)] exhibits a clean symmetric doublet with a site-independent separation, indicating that the second-harmonic Bloch-oscillation frequency is pinned while the fundamental frequencies vary with the injection position.

The site-dependent oscillation periods and spectral structures described above originate from the parity-dependent effective tilts imposed by the quadratic gradient.
The effective potential $V_{\rm{eff,2}} \left( i \right) = (-1)^{i+1} B_{\rm{z,2}} \times i$ grows linearly with site index while reversing sign between sublattices, shearing the Wannier-Stark ladder into two branches with distinct slopes.
In the weak-field limit ($B_{\mathrm{z},2} \ll g$), treating $V_{\rm{eff,2}}$ as a perturbation yields the renormalized tilts
\begin{equation}
    \begin{aligned}
        k_{\rm{eff}} = B_{\rm{z,2}} \mathcal{J}_0 \left( \frac{4t}{g} \right), 
        \end{aligned}
    \label{eq11}
\end{equation}
where the Bessel function $\mathcal{J}_0\left( \frac{4t}{g} \right)$ encodes the non-local response of the doublon defect, averaging the site-dependent potential over the spatial extent $\sim t/g$ of the Wannier-Stark wavefunctions. 
For $t/g\to 0$, the wavefunction collapses to a single site, $\mathcal{J}_0\to 1$, 
and $k_{\rm eff}\to B_{z,2}$. 
At finite $t/g$, the doublon defect probes neighboring sites before resolving the sign alternation, partially washing out the sawtooth and suppressing the shear.

The spectrum thereby organizes into two intertwined branches:
\begin{equation}
    \begin{aligned}
        E_{2n} &\approx (g - k_{\text{eff}}) 2n + E_0, \\
        E_{2n+1} &\approx (g + k_{\text{eff}}) (2n+1) + E_0.
        \end{aligned}
    \label{eq12}
\end{equation}
where the odd branch steepens to $\sim g+k_{\rm{eff}}$ and the even branch flattens to $\sim g-k_{\rm{eff}}$, producing a scissor-like shear.
Consequently, the nearest-neighbor gaps grow linearly with the quantum number $n$,  given by
\begin{equation}
    \begin{aligned}
        &\Delta E_{1}^{+} \left( n \right) =E_{2n+1}-E_{2n} =g+ k_{\rm{eff}} \left( 4n + 1 \right), \\
        &\Delta E_{1}^{-} \left( n \right) =E_{2n+2}-E_{2n+1} =g-k_{\rm{eff}} \left( 4n + 3 \right),
    \end{aligned}
    \label{eq13}
\end{equation}
The dotted lines in Fig.~\ref{fig4}(b) mark the theoretical positions of these nearest-neighbor gaps, showing agreement with the numerical peaks. 
These site-dependent gaps constitute the dominant frequency components of the dynamical evolution; the quantum interference between the overlapping Wannier-Stark states produces the characteristic beating pattern with modulation period $T_{\rm beat}\propto 1/|\Delta E_1^+(n)-\Delta E_1^-(n)|\propto 1/(2n+1)$, directly encoding the sheared ladder structure into the temporal oscillations of Fig.~\ref{fig4}(a).

The next-nearest-neighbor gaps remain $n$-independent:
\begin{equation}
    \begin{aligned}
        &\Delta E_{2}^{+}  =E_{2n+1}-E_{2n-1}   =2 \left( g + k_{\rm{eff}} \right), \\
        &\Delta E_{2}^{-}  =E_{2n+2}-E_{2n}     =2 \left( g - k_{\rm{eff}} \right).
    \end{aligned}
    \label{eq14}
\end{equation}
arising from the fact that states within the same subladder share identical slopes.
The dashed lines in Fig.~\ref{fig4}(c) indicate the theoretical predictions from Eq.~\eqref{eq14}, confirming the fixed doublet separation $4k_{\rm eff}$ for all initial sites and pinning the second-harmonic Bloch-oscillation frequency.

Higher-order corrections refine this scissor geometry: a second-order term $ k_{\rm{c}}=B_{\rm{z,2}}^2[1-\mathcal{J}_0^2(4t/g)]/g$ uniformly shifts both subladder tilts without altering the opening angle, while third-order contributions bend the blades at large excitation numbers ($n\gtrsim 10$), capturing the asymptotic curvature of the spectral fan (Appendix~\ref{app-C}).

This contrasts with the linear case: there, even peaks remain pinned at $2ng$ because intrasubladder transitions feel only the uniform tilt $g$. Under the quadratic gradient, the subladders themselves have distinct slopes $g\pm k_{\rm eff}$, so even peaks split by $4k_{\rm eff}$ and directly encode the shear. This reversal of roles—odd peaks split by $B_{z,1}$ but not by $B_{z,2}$, even peaks the opposite—provides a clean criterion for distinguishing the two gradients.

\subsection{Defect GBO under both linear and quadratic magnetic field}
\label{section:III-C}
\begin{figure}[t]
	\centering
	\includegraphics[trim=10 30 -10 -10, width=0.5 \textwidth]{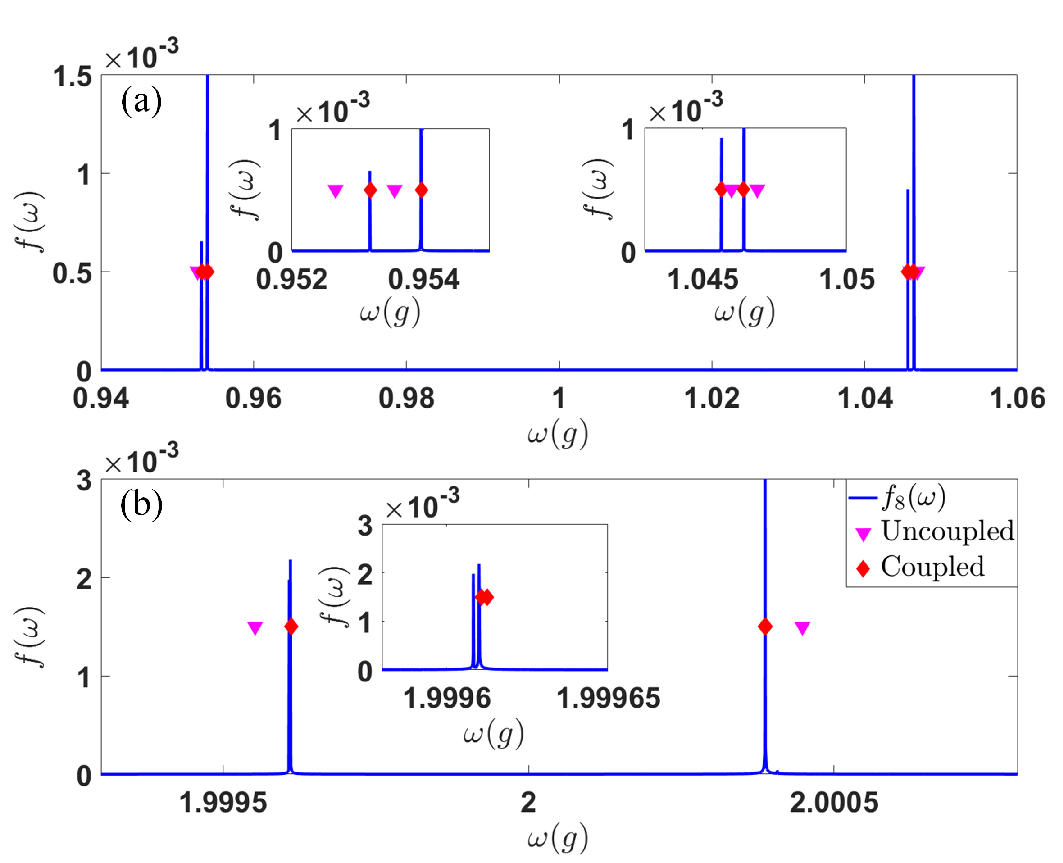}
	\caption{
    \label{fig5}
    Fourier spectrum under simultaneous linear and quadratic magnetic gradients. 
    (a) Spectrum around the fundamental frequency $\omega\approx g$.
    (b) Spectrum around the first harmonic $\omega\approx 2g$.
    In both panels, the blue solid line is the exact numerical result for $f_8(\omega)$.
    The magenta solid triangles and red solid diamonds show the prediction of the uncoupled (na\"ive additive) and coupled theory, respectively.
    Parameters: $N=16$, $g=1$, $t=0.5g$, $B_{z,1}=0.2g$, $B_{z,2}=0.001g$.
    }
\end{figure}

We now turn to the simultaneous presence of both gradients. 
Under a naive additive scheme, the energy levels would be given by $E_n^{\rm (add)}=ng+(-1)^{n+1}(\delta_{\rm weak}+nk_{\rm eff})$, with the sublattice offset $\delta_{\rm weak}$ obtained from Sec.~\ref{section:III-A} and the effective tilt $k_{\rm eff}$ derived from Sec.~\ref{section:III-B}. 
The nearest-neighbor and next-nearest-neighbor gaps then read $\Delta E_1^{\pm}=g\pm\delta_{\rm weak}$ and $\Delta E_2^{\pm}=2(g\pm k_{\rm eff})$, respectively. 
However, as shown in Fig.~\ref{fig5}(a) and \ref{fig5}(b), the spectral peak positions obtained from numerical calculations (blue solid line) exhibit significant deviations from these simple predictions (magenta triangles).
These discrepancies, reaching $\sim 2\%$ relative error near $\omega\approx g$ and comparable magnitude near $\omega\approx 2g$, far exceed the expectations of higher-order perturbative corrections in either isolated limit.
This indicates that the two gradients generate non-additive coupling through the nonlocal quantum response of the doublon, leaving distinct fingerprints on the Bloch-oscillation spectrum: a rigid baseline shift of the fundamental doublet and an asymmetric broadening of the first harmonic.

To quantitatively characterize the non-additive coupling, we treat the total perturbation $V_{\rm eff}=V_{\rm eff,1}+V_{\rm eff,2}$ as a whole and carry out the perturbative calculation up to third order. 
The total energy decomposes into uncoupled and coupled contributions,
\begin{equation}
    E_n=E_n^{\rm (unc)}+E_n^{\rm (cou)}.
    \label{eq15}
\end{equation}
Explicitly,
\begin{subequations}
\begin{align}
E_n^{\text{(unc)}} &= ng + (-1)^{n+1}\delta_{\text{weak}} + (-1)^{n+1}nk_{\text{eff}} \nonumber\\
&\quad + \mathcal{C}_{1}\frac{B_{z,2}^2}{g}n + \mathcal{C}_{2}\frac{B_{z,2}^3}{g^2}(-1)^{n+1},  \label{eq16a}  \\
E_n^{\text{(cou)}} &= \mathcal{C}_{3}\frac{B_{z,1}B_{z,2}}{g} + \mathcal{C}_{4}\frac{B_{z,1}^2B_{z,2}}{g^2}n(-1)^{n+1} \nonumber\\
&\quad + \mathcal{C}_{5}\frac{B_{z,1}B_{z,2}^2}{g^2}(-1)^{n+1},
\label{eq16b}
\end{align}
\end{subequations}
where $\mathcal{C}_{1}$--$\mathcal{C}_{5}$ are dimensionless 
coefficients depending on $\alpha=4t/g$ (see Appendix~\ref{app-D}).

Taking the parameters of Fig.~\ref{fig5} ($B_{z,1}=0.2g$, $B_{z,2}=0.001g$, $t=0.5g$) as a representative weak-field point, the uncoupled terms are dominated by first-order shifts $\delta_{\rm weak}\sim 4 \times 10^{-2}g$ and $nk_{\rm eff}\sim 10^{-3}g$, while pure higher-order contributions ($\sim 10^{-6}g$, $\sim 10^{-9}g$) are negligible.
The coupled terms obey a pronounced hierarchy: the second-order cross-term $C_3 B_{z,1}B_{z,2}/g\sim 10^{-4}g$ is the leading nonlinear correction, and the third-order cross-term $C_4 B_{z,1}^2 B_{z,2}n/g^2\sim 10^{-5}g$ remains non-negligible because its coefficient $C_4$ is large enough to offset the extra suppression $B_{z,1}/g\approx 0.2$. 
Omitting $E_n^{\rm (cou)}$ yields energy-level errors of order $10^{-5}g$; the complete coupled theory improves this by more than a factor of $30$, to $\sim 10^{-7}g$. 
The relative strengths scale as $|E^{(2,\rm cross)}|/\delta_{\rm weak}\sim B_{z,2}/g$ and $|E^{(3,\rm cross)}|/|E^{(2,\rm cross)}|\sim B_{z,1}/g$, with proportionality constants fixed solely by $\alpha=4t/g$; non-additive effects thus grow algebraically and become comparable to first-order shifts when $B_{z,2}\sim g$ or $B_{z,1}\sim g$, signaling the breakdown of the weak-field perturbative regime.

These cross-corrections reshape the generalized Bloch-oscillation spectrum by distorting the underlying scissor geometry in three ways. 
First, the second-order pure term $\mathcal{C}_1 B_{z,2}^2/g$ uniformly shifts both subladder tilts, sliding the scissors without altering the opening angle and producing a global correction to the oscillation frequencies. 
Second, the second-order cross-term $\mathcal{C}_3 B_{z,1}B_{z,2}/g$ induces a global energy displacement that eliminates the mirror symmetry between the two blades, leaving the rigid baseline shift of the fundamental doublet visible in Fig.~\ref{fig5}(a). 
Third, the third-order cross-term $\mathcal{C}_4 B_{z,1}^2 B_{z,2}n(-1)^{n+1}/g^2$ differentially renormalizes the two slopes, changing the scissor opening and generating 
the asymmetric broadening of the first harmonic in Fig.~\ref{fig5}(b); meanwhile, the $\mathcal{C}_5^{(2)}$ contribution introduces a progressive $n^2$ warping at large excitation numbers.

The leakage analysis in Appendices~\ref{app-E} and~\ref{app-F} further supports the intrinsic origin of these spectral fingerprints.
For the parameters of Fig.~\ref{fig5}, the characteristic leakage-induced corrections to the Bloch-oscillation peaks are the staggered offset $|\delta_{\rm L}|\simeq 5.5 \times10^{-9}g$ and the tilt shift $|\gamma_{\rm L}|\simeq 6.4 \times10^{-12}g$.
These shifts are roughly three to four orders of magnitude smaller than the leading nonadditive cross-coupling corrections of order $10^{-5}g$--$10^{-4}g$.
The rigid baseline shift of the fundamental doublet and the asymmetric broadening of the first harmonic therefore arise from the coupled magnetic gradients, rather than from virtual leakage outside the truncated Hilbert space.

Quantitatively, the coupled scheme reduces the energy-level error from $\sim 10^{-5}g$ to $\sim 10^{-7}g$, confirming that resolving the fine structure of the defect's Bloch-oscillation spectrum requires the inclusion of cross-terms. 
The non-additive coupling illustrates a general principle: distinct field channels in a correlated system do not respond independently but interfere through shared virtual processes. 
A two-site selection rule that forbids certain responses in one channel can be broken by another,
allowing cross-terms to dominate over the higher-order pure terms of the 
weaker channel. 
This gradient-induced unlocking mechanism suggests that multi-gradient spectroscopy of defects can probe parity-protected selection rules and their controlled breaking in quantum magnets.

\section{Experimental feasibility}
\label{section:IV}

The model Hamiltonian can be realized with fermionic magnetic atoms ($^{161}$Dy or $^{167}$Er)~\cite{Lu2012PRL,Aikawa2014PRL,Chomaz2023RPP} in a deep one-dimensional optical lattice, with spin-$1/2$ encoded in the stretched Zeeman states $|m_F=\pm F\rangle$ and dipoles polarized perpendicular to the chain. 
The dipole--dipole interaction then provides an antiferromagnetic Ising coupling $J_z/h\sim1$--$2~\mathrm{kHz}$ at a spacing $d\simeq130$--$140~\mathrm{nm}$, while both first-order and residual dipolar exchange are suppressed by the angular-momentum constraint $|\Delta m_F|=2F\gg2$ and by quadratic-Zeeman detuning under tight confinement, respectively~\cite{Chomaz2023RPP, Barral2024NC}.
The on-site interaction $U$ is tunable via Feshbach resonances~\cite{Maier2015PRA, Frisch2014Nature, FerrierBarbut2016PRL}, and microwave-shielded polar molecules~\cite{Anderegg2021Science, Schindewolf2022Nature} or Rydberg-dressed lattice gases~\cite{Zeiher2016NPHYS, Zeiher2017PRX, Montorsi2020PRA} offer alternative realizations. 
A representative parameter set $J_z/h\simeq2~\mathrm{kHz}$, $U/h\simeq4~\mathrm{kHz}$ ($U=2J_z=200\,g$), $t/h\simeq10~\mathrm{Hz}$, and $g/h\simeq20~\mathrm{Hz}$ yields a Bloch period of $50~\mathrm{ms}$. 
The spin-independent tilt $g$ can be supplied by gravity with the lattice inclined by $\sim0.2^{\circ}$ from the horizontal, or by an optical potential gradient. 
The linear magnetic gradient $B_{z,1}$ requires only $\sim0.1~\mathrm{G/cm}$ for the large dysprosium moment~\cite{dePaz2013PRL, Geiger2018PRL}, and the quadratic term $B_{z,2}$ follows from magnetic-field curvature, which couples directly to the opposite moments of the stretched-state pair. 
The initial N\'eel-ordered chain with a single doublon is a product state, where the spin order can be assembled by superlattice protocols~\cite{Trotzky2008Science, Boll2016Science, Mazurenko2017Nature} and single-doublon realizations can be post-selected from parity-projecting quantum-gas-microscope images, which also provide dynamical readout~\cite{Bakr2009Nature, Sherson2010Nature}.

Coherent spin dynamics over $10$--$100~\mathrm{ms}$ has been demonstrated in dipolar lattices~\cite{dePaz2013PRL, Yan2013Nature, Patscheider2020PRR}, and Bloch oscillations of weakly interacting condensates have been sustained for seconds~\cite{Ferrari2006PRL}, while extending such coherence to interacting lattice spin systems remains the principal challenge.
The doublet splitting $\delta_{\rm weak}\simeq \mathcal{J}_0(4t/g)B_{z,1}$ ranges from $\sim1~\mathrm{Hz}$ at $B_{z,1}=0.2g$ to $\sim10$--$20~\mathrm{Hz}$ at $B_{z,1}=2$--$5g$ (Fig.~\ref{fig3}), with the upper end resolvable within current coherence times. 
Verifying the shearing and cross-coupling at the parameters of Fig.~\ref{fig5} ($10^{-4}$--$10^{-7}g$ features) would require $10^4$ or more coherent Bloch periods, but enhanced gradients ($B_{z,1}\sim g$, $B_{z,2}\sim10^{-2}g$) bring the same signatures into the $0.1$--$1~\mathrm{Hz}$ range, accessible as coherence times approach the second scale. 
Leakage imposes no practical limitation, since at $t/\Delta_{\rm leak}=1/500$ the associated peak shifts are $\sim5\times10^{-9}g$ (Appendix~\ref{app-F}), far below all scales discussed above.\\

\section{Conclusion }
\label{section:V}
We have investigated GBOs of a doublon defect in a one-dimensional antiferromagnetic $t$-$J_z$ chain under simultaneous linear and quadratic magnetic field gradients. 
Mapping the many-body dynamics onto an effective single-particle model reveals how the N\'eel background imprints parity-dependent structures on the defect's periodic motion:
the linear gradient generates a characteristic beat pattern via Wannier-Stark subladder splitting with offset $\delta$; 
the quadratic gradient produces site-dependent oscillation periods by shearing the ladder into a scissor-like profile with renormalized slope $k_{\rm eff}=B_{z,2}\mathcal{J}_0(4t/g)$. 
When both gradients coexist, their interplay is fundamentally nonadditive. 
The quadratic gradient breaks the two-site periodicity that forbids second-order linear responses, unlocking a dominant cross-coupling $\propto B_{z,1}B_{z,2}$ that biases the fundamental doublet and produces asymmetric broadening of the first harmonic---spectral fingerprints invisible to any uncoupled theory and quantitatively corroborated by exact many-body simulations.

The model is within reach of current quantum-simulation platforms. 
As detailed in the experimental-feasibility discussion above, fermionic magnetic atoms ($^{161}$Dy or $^{167}$Er) in a deep optical lattice provide the antiferromagnetic Ising coupling $J_z/h\sim1$--$2~\mathrm{kHz}$ directly from dipole--dipole interactions between stretched Zeeman states, with the on-site interaction tunable via Feshbach resonances; 
Rydberg-dressed lattice gases~\cite{Zeiher2016NPHYS,Zeiher2017PRX,Montorsi2020PRA} and microwave-shielded polar molecules~\cite{Anderegg2021Science,Schindewolf2022Nature} offer alternative realizations. 
The tilt $g$ can be supplied by gravity or a spin-independent optical gradient, the linear gradient $B_{z,1}$ requires magnetic gradients of only $\sim0.1~\mathrm{G/cm}$ for the large dysprosium moment, and the quadratic gradient $B_{z,2}$ is a magnetic-field curvature that couples directly to the opposite moments of the stretched-state pair.
The full parameter hierarchy $J_z,U\gg t,g\gg B_{z,1}\gg B_{z,2}$ is therefore compatible with demonstrated experimental capabilities, and leakage out of the defect subspace remains negligible ($t/\Delta_{\rm leak}=1/500$ at our working point; Appendix~\ref{app-F}).

Our work establishes a gradient-induced unlocking mechanism for multi-gradient defect spectroscopy, offering a framework for probing parity-protected selection rules and their controlled breaking in tunable quantum simulators.

\section*{Acknowledgements}
This work was supported by the High Level Talents Research and Startup Foundation Projects for Doctors of Zhoukou Normal University (No.ZKNUC2023016),  the Natural Science Foundation Project of Henan Province of China (No.22300420941, No.252300423541, No.262300422459), the Research Project of Hubei Education Department(No. B2023077) and the Doctoral Scientific Research Foundation of Hubei University of Automotive Technology (No.BK202432).

\begin{appendix}

\section{Second-order leakage corrections to the effective single-particle description}
\label{app-E}

\setcounter{equation}{0}
\renewcommand{\theequation}{A\arabic{equation}} 

In this appendix, we assess the validity of the effective single-particle description 
[Eq.~\ref{eq3} of the main text] by estimating corrections arising from leakage out of the truncated Hilbert space $\mathcal{H}=\{ \left| i \right\rangle \}$. 
Such leakage arises from single-atom hopping processes that transiently create an extra hole-doublon (HD) or doublon-hole (DH) pair in the Néel background, and consequently yields corrections to Eq.~\ref{eq3} of the main text.
To estimate these corrections, we decompose the Hamiltonian in Eq.~\ref{eq1} as
\begin{equation}
    \begin{aligned}
    \hat{H}=\hat{H}_0+\hat{H}_t,
    \end{aligned}
\label{eqSE1}
\end{equation}
where $\hat{H}_0$ contains all diagonal terms and 
\begin{equation}
    \begin{aligned}
    \hat{H}_0 & =  J_{z} \sum\noindent_{ \langle i,j \rangle } ^N\hat{S}_i^z \hat{S}_j^z +  U \sum\noindent_i^N \hat n_{i,\uparrow}\hat n_{i,\downarrow} + \sum\noindent_{i=1}^{N} \left( g \times i \right) \hat{n}_{i} \\
    & + \sum\noindent_{i=1}^{N} \left( B_{\rm{z,1}} \times i + B_{\rm{z,2}} \times i^2 \right) \hat{S}_{i}^z,
    \end{aligned}
\label{eqSE2}
\end{equation}
and the hopping term $\hat{H}_t$ is treated as the perturbation
\begin{equation}
    \begin{aligned}
    \hat{H}_t=&-t\sum\noindent_{\langle i,j \rangle }^{N} \sum\noindent_{\sigma=\uparrow,\downarrow} \hat{a}_{i,\sigma}^{\dagger} \hat{a}_{j,\sigma}.
    \end{aligned}
\label{eqSE3}
\end{equation}

\begin{figure}[t]
	\centering
	\includegraphics[trim=0 20 -30 -10, width=0.5 \textwidth]{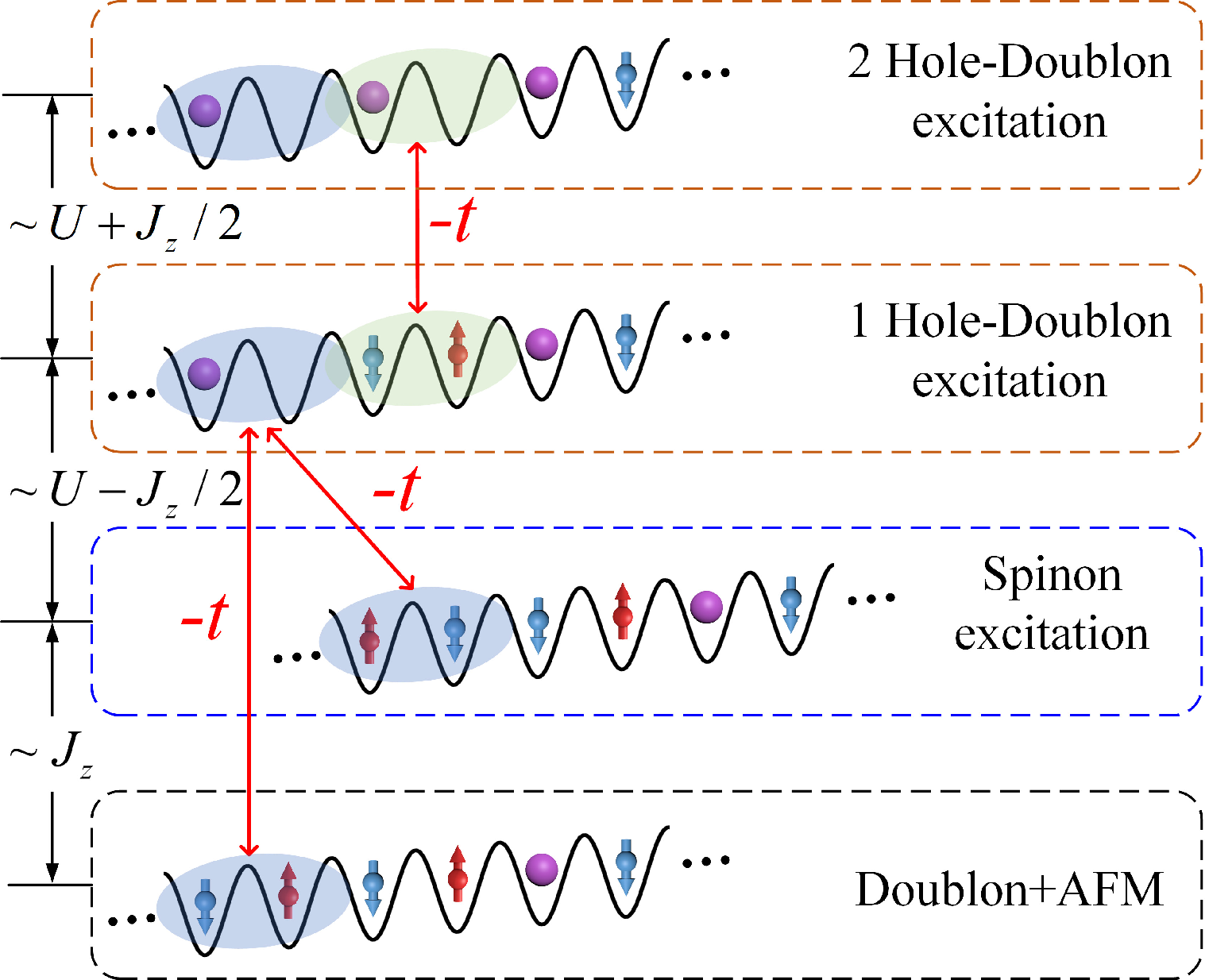}
	\caption{
    \label{figE1}
    Leakage channels out of the defect subspace, ordered by excitation energy (not to scale). 
    A single hop ($-t$, red arrows) takes a defect state $|i\rangle$ (bottom) to a hole--doublon (HD) intermediate state at $\tilde{\Delta}_{\rm leak}\sim U+J_z/2$; a second hop reaches a second HD pair or a spinon excitation ($\sim J_z$), contributing at order $t^4/\Delta_{\rm leak}^3$. 
    The spinon state, though energetically lower, is not accessible by any single hop.
    }
\end{figure}

Starting from a basis state $|i\rangle\in\mathcal{H}$, a single hop in $\hat{H}_t$ either moves the doublon to a neighboring site (thereby staying inside $\mathcal{H}$) or creates an extra hole--doublon (HD) pair on a bond $\langle j,j+1\rangle$ with $j\neq i-1,i$, as summarized in the level diagram of Fig.~\ref{figE1}. 
No single hop can create a spinon from $|i\rangle$, because $\hat{H}_t$ moves atoms without flipping their spins and thus cannot break a N\'eel pair without simultaneously creating an HD excitation. 
Spinon excitations, although energetically cheaper ($\sim J_z$) than HD pairs, are therefore reached only through an HD intermediate state via a second hop [Fig.~\ref{figE1}], contributing to the energy at order $t^4/\Delta_{\rm leak}^3$ and negligible in the present analysis.

For N\'eel bond $\langle j,j+1\rangle$ where an HD or DH pair can be created, the spins are antiparallel, $s_{j+1}=-s_j$ with $s_j=\pm 1/2$.
Each such bond hosts two leakage channels: an atom hopping to the right creates an HD pair, while hopping to the left creates a DH pair.
The energy denominators for these virtual processes are
\begin{equation}
    \begin{aligned}
    \Delta E_{\pm}(j)=\tilde{\Delta}_{{\rm{leak}}} (j)  \pm g+s_j\Delta B_j, 
    \end{aligned}
    \label{eqSE4}
\end{equation}
where $\tilde{\Delta}_{{\rm{leak}}} (j) =U+\nu_j J_z/4$ contains the on-site energy of the extra doublon and the cost of broken Ising bonds, and $\Delta B_j=B_{z,1}+B_{z,2}(2j+1)$ is the Zeeman energy difference across the bond. 
The $+$ ($-$) sign refers to the final state $|HD\rangle$ ($|DH\rangle$) in which the atom hops to the right (left). 
The Zeeman change  $s_j\Delta B_j=-(s_jB_j+s_{j+1}B_{j+1})$ with $B_j\equiv B_{z,1}j+B_{z,2}j^2$ reflects the staggered spin polarization of the N\'eel background.
The broken-bond count $\nu_{j}$ depends on the position of the HD pair relative to the original doublon.
For bonds adjacent to the defect [$(i-2,i-1)$ or $(i+1,i+2)$], as well as the end bonds $(1,2)$ and $(N-1,N)$, two Ising bonds are broken ($\nu_{j}=2$).
For all other internal bonds the HD pair sits in the bulk of the N\'eel background and three Ising bonds are broken ($\nu_{j}=3$). 
These two cases define the characteristic energy gaps
\begin{equation}
    \begin{aligned}
    \Delta_{\rm{leak}}\equiv U+\frac{J_{z}}{2}, \qquad 
    \Delta_{\rm{leak}}' \equiv U+\frac{3J_{z}}{4},
    \end{aligned}
    \label{eqSE5}
\end{equation}
corresponding to $\nu_j=2$ and $\nu_j=3$, respectively.

Substituting Eq.~(\ref{eqSE4}) into the standard second-order perturbation formula, with $|\langle n|\hat{H}_t|i\rangle|^2=t^2$ for each allowed hop, we obtain
\begin{equation}
    E^{(2)}_i=-t^{2}\sum_{j\notin\{i-1,i\}}\left[\frac{1}{\tilde{\Delta}_{\rm{leak}} (j) +g+s_{j} {\Delta} B_{j}}+\frac{1}{\tilde{\Delta}_{\rm{leak}} (j) -g+s_{j}{\Delta} B_{j}}\right],
\label{eqSE6}
\end{equation}
valid for any chain length $N$ and any defect site $i$. The excluded bonds $j=i-1,i$ correspond to hops onto the doublon site, which are Pauli forbidden.

In the weak-field limit $g,B_{z,1},B_{z,2}\ll\Delta_{\rm{leak}},\Delta_{\rm{leak}}'$, we expand Eq.~(\ref{eqSE6}) to linear order in $B_{z,1}$ and $B_{z,2}$ and to bilinear order in $B_{z,1}B_{z,2}$. 
For a bulk site $4\le i\le N-3$, the remaining $N-3$ bonds separate into four bonds with $\nu_j=2$ [$j=1,i-2,i+1,N-1$] and $N-7$ bonds with $\nu_j=3$. 
Performing the staggered lattice sums over $ s_j=\pm1/2$, the second-order shift takes the form
\begin{equation}
    \begin{aligned}
    E^{(2)}_i & =\sigma_i\,\kappa_{\mathrm{eff}}\,B_{z,2} 
     +\bigl(A_{\mathrm{cross}}+
    B_{\mathrm{cross}}\cdot i\bigr)B_{z,1}B_{z,2} \\
    &+C^{(2)}+ \mathcal{O}(B_{z,1}^2,B_{z,2}^2,g^2),
    \end{aligned}
    \label{eqSE7}
\end{equation}
The absence of site-dependent term linear in $B_{z,1}$ follows because the staggered sums $\sum s_j$ for both $\nu_j=2$ and $\nu_j=3$ bonds are independent of the defect site $i$. 
The $B_{z,2}$ response is a pure staggered shift proportional to $\sigma_i$, with no spurious linear-$i$ factor; its renormalized coefficient is
\begin{equation}
    \begin{aligned}
    \kappa_{\mathrm{eff}}\equiv t^2[-3w_{\Delta_{\rm{leak}}}+2w_{\Delta_{\rm{leak}}'}],
    \end{aligned}
    \label{eqSE8}
\end{equation}
where $w_X\equiv 1/(X+g)^2+1/(X-g)^2$.
The $B_{z,1}B_{z,2}$ cross term carries a linear-$i$ dependence with slope
\begin{equation}
    B_{\mathrm{cross}} = -2t^2[w_{\Delta_{\rm{leak}}}^{(3)} -2w_{\Delta_{\rm{leak}}'}^{(3)}],
\label{eqSE9}
\end{equation}
where $w_X^{(3)}\equiv 1/(X+g)^3+1/(X-g)^3$.
And the $i$-independent offset
\begin{equation}
    A_{\mathrm{cross}}=-\frac{t^2}{2}\left[(2N+2)w_{\Delta_{\rm{leak}}}^{(3)}+(N^2-2N-3)w_{\Delta_{\rm{leak}}'}^{(3)}\right]
\label{eqSE10}
\end{equation}
is absorbed into the uniform shift $E_0$ of the main text. 
The field-independent constant
\begin{equation}
    C^{(2)}=-2t^2\left[\frac{4\Delta_{\rm{leak}}}{\Delta_{\rm{leak}}^2-g^2}+\frac{(N-7)\Delta_{\rm{leak}}'}{(\Delta_{\rm{leak}}')^2-g^2}\right]
\label{eqSE11}
\end{equation}
collects all zero-field contributions.

\begin{figure}[t]
	\centering
	\includegraphics[trim=0 20 -20 -10, width=0.5 \textwidth]{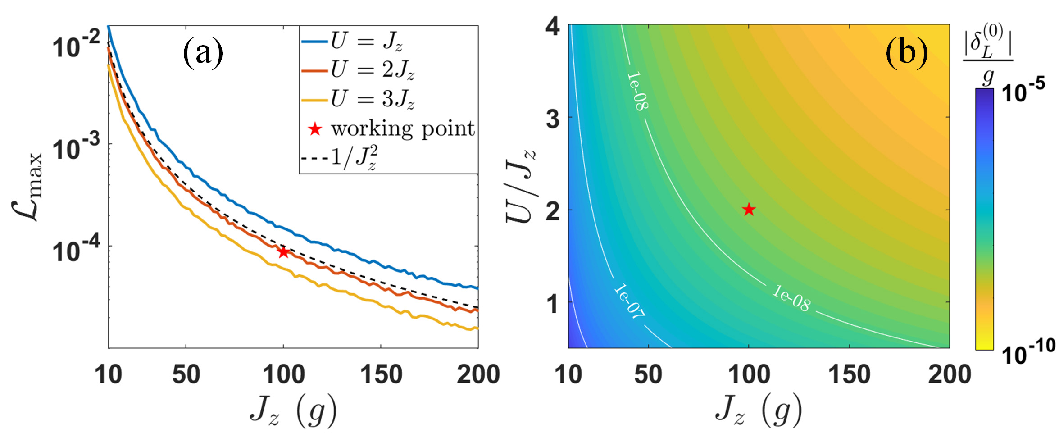}
	\caption{
    \label{figE2}
    (a) Maximum leakage weight $\mathcal{L}_{\max}$ [Eq.~(\ref{eqSE12})] out of the defect subspace versus $J_z$, obtained from exact time evolution of the full Hamiltonian Eq.~(\ref{eq1}) over $10^4$ Bloch periods, for $U=J_z$, $2J_z$, and $3J_z$ (top to bottom). The dashed line indicates the scaling $\propto J_z^{-2}$.
    (b) Leakage-induced peak shift $|\delta_L^{(0)}|/g$ in the $(J_z,U)$ plane, evaluated from Eq.~(\ref{eqSE8}) at $B_{z,2}=10^{-3}g$; white lines mark the $10^{-7}$ and $10^{-8}$ contours.
    In both panels the red star marks the working point ($J_z=100g$, $U=200g$). Other parameters: $N=16$, $t=0.5g$, $B_{z,1}=0.2g$.
    }
\end{figure}

The perturbative analysis above rests on the small parameter $t/\Delta_{\rm leak}$, and its validity is most directly tested by the leakage weight itself. 
We therefore perform exact time evolution of the full microscopic Hamiltonian Eq.~(\ref{eq1}), without any truncation, starting from the initial state $|i=8\rangle$ of the main text, and compute the total weight of the wave function outside the defect subspace,
\begin{equation}
    \mathcal{L}(t)=1-\sum_{i=1}^{N}\left|\langle i|\psi(t)\rangle\right|^{2},
    \label{eqSE12}
\end{equation}
with $\mathcal{L}_{\max}$ denoting its maximum over $10^4$ Bloch periods. 
As shown in Fig.~\ref{figE2}(a), $\mathcal{L}_{\max}$ follows the perturbative scaling $(t/\Delta_{\rm leak})^{2}\propto J_z^{-2}$ for $U/J_z=1,2,3$, confirming that leakage is exhausted by the second-order virtual processes enumerated above; the upturn at $J_z/g\lesssim20$ marks the breakdown boundary of the truncated description. 
At the working point ($J_z=100g$, $U=200g$), $\mathcal{L}_{\max}=8.7\times10^{-5}$, i.e., the wave function retains over $99.99\%$ of its weight inside $\mathcal{H}$ throughout the evolution.

The spectral positions, on which all our conclusions rest, are far less sensitive than the leakage weight itself. 
The level shift is set by $\kappa_{\rm eff}$ [Eq.~(\ref{eqSE8})], in which the two virtual-state families nearly cancel, and is further compressed by the linear-response factor $\mathcal{J}_0(4t/g)\approx0.22$. 
As mapped in Fig.~\ref{figE2}(b), the resulting peak shift $|\delta_L^{(0)}|$ decreases monotonically with both $J_z$ and $U$, remains below $10^{-7}g$ for $J_z\gtrsim 22g$ [Fig.~\ref{figE2}(b)], and reaches only $\sim5\times10^{-9}g$ at the working point---four orders of magnitude below $\mathcal{L}_{\max}$ and well below the weakest spectral feature in Fig.~\ref{fig5}. 
Such a shift is unresolvable to exact diagonalization and is therefore evaluated analytically, while the truncation itself is benchmarked by the measurable $\mathcal{L}_{\max}$. 
The leakage weight affects only the spectral contrast, at the same $10^{-4}$ relative level. 
Leakage is thus negligible on all three levels: the Hilbert-space truncation, the peak positions, and the spectral weights.\\

\section{Perturbative calculation of $\delta_{\rm{weak}}$ in the weak-field limit}
\label{app-A}
\setcounter{equation}{0}
\renewcommand{\theequation}{B\arabic{equation}} 
In this appendix, we derive the $\delta$ within a theoretical third-order perturbation treatment based on the effective single-body Hamiltonian Eq.~\ref{eq3}.
For the case where $B_{\rm{z},1} \ll t,g $ and $B_{\rm{z},2}=0$, that is, when only a linear magnetic gradient is applied to the spin chain, the Hamiltonian in Eq.~\ref{eq3} can be written as
\begin{equation}
    \hat{H}_{\rm{eff}} =\hat{H}_0 + V_{\rm{eff,1}},
    \label{eqSA1}
\end{equation}
where $\hat{H}_0$ is the leading term and 
\begin{equation}
    \hat{H}_0 =-t \sum\noindent_{i=1}^N \left( \left| i \right\rangle \left\langle i+1 \right| +  \left| i+1 \right\rangle \left\langle i \right|   \right)  \\
    +g \sum\noindent_{i=1}^N i \left| i \right\rangle \left\langle i \right| ,
    \label{eqSA2}
\end{equation}
and the perturbation of the staggered potential is
\begin{equation}
    V_{\rm{eff,1}}= \frac{B_{\rm{z},1}}{2} \sum\noindent_{i=1}^N \left( -1 \right)^{i+1}  \left| i \right\rangle \left\langle i \right| .
    \label{eqSA3}
\end{equation}

The eigenenergy (to zeroth order) and eigenstates of $\hat{H}_0$ is the Wannier-Stark ladder $E^{(0)}_n=ng $ and Wannier-Stark states 
\begin{equation}
    \left|  {\psi_n ^{(0)} }  \right\rangle = \sum_i \mathcal{J}_{i-n} \left( \frac{2t}{g} \right) \left|  i  \right\rangle ,
    \label{eqSA4}
\end{equation}
where $\mathcal{J}_{i-n}$ is the $i-n$th order Bessel function of the first kind. 

In the basis $\{ \left| \psi_n^{(0)} \right\rangle \}$, the matrix elements of perturbation are
\begin{equation}
    \begin{aligned}
        V_{n,n+p}
        &=\left\langle {\psi_n ^{(0)} } \left|   V_{\rm{eff,1}}  \right| {\psi_{n+p} ^{(0)} } \right\rangle \\
        &= \frac{B_{\rm{z},1}}{2}  \sum_i^N \mathcal{J}_{i-n}  (\frac{2t}{g}) \mathcal{J}_{i-n-p} (\frac{2t}{g}) \left( -1 \right)^{i+1},
    \end{aligned}
    \label{eqSA5}
\end{equation}
In the thermodynamic limit \(N \to \infty\) and let $k=i-n$, then the summation becomes 
\begin{equation}
    \begin{aligned}
        V_{n,n+p} = \frac{B_{\rm{z},1}}{2} \sum_{k=-\infty}^{\infty} \mathcal{J}_k(\frac{2t}{g}) \mathcal{J}_{k-p}(\frac{2t}{g}) (-1)^k ,
    \end{aligned}
    \label{eqSA6}
\end{equation}
Using the identity $(-1)^k\mathcal{J}_k(\frac{2t}{g})=\mathcal{J}_k(-\frac{2t}{g})$ and the Bessel addition theorem $\sum\noindent_{k=-\infty}^\infty \mathcal{J}_k (x) \mathcal{J}_{k+d} (y) = \mathcal{J}_{d} (x-y) $, we have
\begin{equation}
    \begin{aligned}
        V_{n,n+p}=\frac{B_{\rm{z},1}}{2}  (-1)^{n+1} \mathcal{J}_{-p} (\alpha) ,
    \end{aligned}
    \label{eqSA7}
\end{equation}
where $\alpha=\frac{4t}{g}$.

The first-, second-, and third-order corrections to the eigenenergies are then
\begin{equation}
    \begin{aligned}
        E_n^{(1)} =  V_{n,n} = \frac{B_{\rm{z},1}}{2}   \mathcal{J}_{0} (\alpha) \left( -1 \right)^{n+1},
    \end{aligned}
    \label{eqSA8}
\end{equation}
\begin{equation}
    \begin{aligned}
        E_n^{(2)}  = \sum_{p \neq 0}^N \frac{ |V_{n,n+p}|^2}{E_n^{(0)}-E_{n+p}^{(0)}} =0,
    \end{aligned}
    \label{eqSA9}
\end{equation}
and
\begin{equation}
    \begin{aligned}
        E_n^{(3)} 
    &= \sideset{}{'}\sum_{p,q} \frac{V_{n,n+p} \, V_{n+p,n+q} \, V_{n+q,n}}{(-pg)(-qg)} \\ 
    & - V_{nn} \sideset{}{'}\sum_{p} \frac{|V_{n,n+p}|^{2}}{(-pg)^{2}} \\
        &= \frac{B_{z,1}^3}{ 8 g^2} (-1)^{n+1} \, \mathcal{C}(\alpha) ,
    \end{aligned}
    \label{eqSA10}
\end{equation}
where $\mathcal{C}(\alpha)$ is the $n$-independent parameter  
\begin{equation}
    \begin{aligned}
        \mathcal{C}(\alpha) &= \sum_{k \neq 0} \sum_{j \neq 0} 
        \frac{(-1)^{k} \, \mathcal{J}_{k}(\alpha) \, \mathcal{J}_{j}(\alpha) \, \mathcal{J}_{j-k}(\alpha)}{k \, j} 
        - \mathcal{J}_0(\alpha) \sum_{k \neq 0} \frac{\mathcal{J}_{k}^2(\alpha)}{k^2} .
    \end{aligned}
    \label{eqSA11}
\end{equation}

Collecting terms up to third order, the eigenenergy becomes
\begin{equation}
    \begin{aligned}
        E_n &= E_n^{(0)}+E_n^{(1)}+E_n^{(3)} \\
        &= ng+  \frac{\left( -1 \right)^{n+1}}{2} \left[ B_{\rm{z},1}   \mathcal{J}_{0} (\alpha)  +\frac{B_{z,1}^3}{g^2}  \mathcal{C}(\alpha) \right] .
    \end{aligned}
    \label{eqSA12}
\end{equation}

The energy gaps between adjacent eigenstates follow as 
\begin{equation}
    \begin{aligned}
        \Delta E_{1}^{\rm{odd \to even}} & =E_{2\ell+1}-E_{2\ell} \\
        & =g+\left[ B_{\rm{z},1}   \mathcal{J}_{0} (\alpha)  +\frac{B_{z,1}^3}{g^2}  \mathcal{C}(\alpha) \right], \\
        \Delta E_{1}^{\rm{even \to odd}} & =E_{2\ell+2}-E_{2\ell+1} \\
        & =g-\left[ B_{\rm{z},1}   \mathcal{J}_{0} (\alpha)  +\frac{B_{z,1}^3}{g^2}  \mathcal{C}(\alpha) \right],
    \end{aligned}
    \label{eqSA13}
\end{equation}
Therefore, the subband offset $\delta_{\rm{weak}}$ in Eq.~\ref{eq8} of the main text, is
\begin{equation}
    \begin{aligned}
        \delta_{\rm{weak}}=B_{\rm{z},1}   \mathcal{J}_{0} (\alpha)  +\frac{B_{z,1}^3}{ 4  g^2}  \mathcal{C}(\alpha)
    \end{aligned}
    \label{eqSA14}
\end{equation}
and we define $\mathcal{A}_3=\frac{1}{4g^2}\mathcal{C}(\frac{4t}{g})$ and $\mathcal{A}_3=\frac{1}{g^2}  \mathcal{C}(\frac{4t}{g})$. 
While the next-nearest-neighbor gap remains
\begin{equation}
    \begin{aligned}
        \Delta E_{2}=2g.
    \end{aligned}
    \label{eqSA15}
\end{equation}

\section{Perturbative calculation of $\delta_{\rm{medium}}$ and $\delta_{\rm{strong}}$ in the strong-field limit}
\label{app-B}
\setcounter{equation}{0}
\renewcommand{\theequation}{C\arabic{equation}} 

In the strong-field limit ($B_{\mathrm{z,1}} \gg t, g$), the staggered potential $V_{\rm{eff,1}} \left( i \right) = (-1)^{i+1} B_{\rm{z,1}}/2$ dominates, imposing a natural two-site periodicity.
We partition the chain into dimers comprising adjacent odd-even sites $\left( 2n-1, 2n \right))$, diagonalize the intradimer Hamiltonian exactly, and treat the inter-dimer tunneling as a perturbation.

The Hamiltonian in Eq.~\ref{eq3} was rewritten as 
\begin{equation}
    \begin{aligned}
        \hat{H} = \sum_n \hat{H}_n + \sum_n \hat{T}_{n,n+1},
    \end{aligned}
    \label{eqSB1}
\end{equation}
where $\hat{H}_n$ is the intradimer Hamiltonian and $\hat{T}_{n,n+1}$ denotes the inter-dimer tunneling. 
Explicitly, 
\begin{equation}
    \begin{aligned}
        \hat{H}_{n} &=  \left[ (2n-1)g + \frac{B_{\mathrm{z},1}}{2} \right] |2n-1\rangle\langle 2n-1| \\
        & + \left[ 2ng - \frac{B_{\mathrm{z},1}}{2}  \right]  |2n\rangle\langle 2n|  \\
        & - t\left(|2n-1\rangle\langle 2n| + |2n\rangle\langle 2n-1|\right),
    \end{aligned}
    \label{eqSB2}
\end{equation}
and
\begin{equation}
    \begin{aligned}
        \hat{T}_{n,n+1} = -t\left(|2n\rangle\langle 2n+1| + |2n+1\rangle\langle 2n|\right).
    \end{aligned}
    \label{eqSB3}
\end{equation}

At zeroth order, the dimers are decoupled.
Diagonalizing the $2 \times 2$ intradimer Hamiltonian $\hat{H}_{n}$ yields two eigenstates per cell: the high-energy (odd-like) state $\left| n,+ \right\rangle$ and the low-energy (even-like) state $\left| n,- \right\rangle$ , given by
\begin{equation}
    \begin{aligned}
        |n,+\rangle &= \cos\theta\,|2n-1\rangle + \sin\theta\,|2n\rangle, \qquad E_{n,+}^{(0)} = \bar{\varepsilon}_{n} + \Delta, \\
        |n,-\rangle &= -\sin\theta\,|2n-1\rangle + \cos\theta\,|2n\rangle, \ \quad E_{n,-}^{(0)} = \bar{\varepsilon}_{n} - \Delta,
    \end{aligned}
    \label{eqSB4}
\end{equation}
where $\bar{\varepsilon}_{n}=(2n-1/2)g$, $\Delta = \sqrt{[(B_{\mathrm{z},1}-g)/2]^2 + t^2}$, and trigonometric functions have the explicit forms:
\begin{equation}
    \begin{aligned}
        \cos\theta = \sqrt{\frac{1}{2} + \frac{B_{\mathrm{z},1}-g}{4\Delta}}, \\
        \sin\theta = \sqrt{\frac{1}{2} - \frac{B_{\mathrm{z},1}-g}{4\Delta}},
    \end{aligned}
    \label{eqSB5}
\end{equation}
When $B_{\mathrm{z},1} = g$, the two sublattices are energetically degenerate and $\theta = \pi/4$. In the strong-field limit ($B_{\mathrm{z},1} \gg g$), one finds $\cos\theta \to 1$ and $\sin\theta \to t/(B_{\mathrm{z},1}-g) \ll 1$, reflecting that the high-energy state $|n,+\rangle$ becomes predominantly localized on the odd site.

In the dimer eigenbasis $\{\left| n, \pm \right\rangle\}$, intradime hopping term $\hat{T}_{n,n+1}$ has the form
\begin{equation}
    \begin{aligned}
    \hat{T}_{n,n+1} = &-t\sin\theta\cos\theta \Big(|n,+\rangle\langle n+1,+| + \mathrm{h.c.}\Big)\\
    &+ t\sin\theta\cos\theta \Big( |n,-\rangle\langle n+1,-| + \mathrm{h.c.} \Big) \\
    &- t\cos^{2}\theta \Big( |n,-\rangle\langle n+1,+| + \mathrm{h.c.} \Big)\\ 
    &+ t\sin^{2}\theta \Big( |n,+ \rangle\langle n+1,- | + \mathrm{h.c.}\Big).
    \end{aligned}
    \label{eqSB6}
\end{equation}

Treating the inter-dimer tunneling $\hat{T} = \sum\noindent_n \hat{T}_{n,n+1}$  as a perturbation to the zeroth-order Hamiltonian $\hat{H}_0 = \sum\noindent_n \hat{H}_{n}$, we expand the eigenenergies as 
\begin{equation}
    E_{n,\pm} = E_{n,\pm}^{(0)} + E_{n,\pm}^{(1)} + E_{n,\pm}^{(2)} + E_{n,\pm}^{(3)} + \cdots.
    \label{eqSB7}
\end{equation}
Owing to the strictly off-diagonal tridiagonal structure of $\hat{T}$, which connects only adjacent dimers $n \leftrightarrow n\pm 1$. 
An odd number of virtual transitions cannot form a closed path returning to the initial dimer state, yielding all odd-order perturbative corrections $E_{n,\pm}^{(1)} = E_{n,\pm}^{(3)} = 0$.

The second-order energy shift involves virtual transitions to the four neighboring states (two left, two right):
\begin{equation}
\begin{aligned}
    E_{n,-}^{(2)} &= \sum_{m,\alpha} \frac{|\langle m,\alpha | \hat{T} | n,- \rangle|^2}{E_{n,-}^{(0)} - E_{m,\alpha}^{(0)}} \\
    &= t^2\left[\frac{\cos^4\theta}{-2(g+\Delta)}  + \frac{\sin^4\theta}{2(g-\Delta)} \right] ,
\end{aligned}
\label{eqSB8}
\end{equation}
where the contributions from the two low-energy channels (proportional to $\sin^2\theta\cos^2\theta$) cancel exactly due to their opposite energy denominators $-2g$ and $+2g$. By symmetry, $E_{n,+}^{(2)} = -E_{n,-}^{(2)}$.

Collecting the perturbative corrections, the eigenenergies of the dimer states are given by
\begin{equation}
\begin{aligned}
    E_{2n-1} &= E_{n,+}^{(0)} + E_{n,+}^{(2)} \\
    &= (2n-\frac{1}{2})g + \left( \Delta - E_{n,-}^{(2)} \right) \\
    &= (2n-1)g + \left( \Delta - E_{n,-}^{(2)}  + \frac{g}{2}\right), \\
    E_{2n} &= E_{n,-}^{(0)} + E_{n,-}^{(2)} \\
    &= (2n-\frac{1}{2})g - \left( \Delta - E_{n,-}^{(2)}  \right) \\
    &= 2n g - \left( \Delta - E_{n,-}^{(2)}  + \frac{g}{2} \right) \\
\end{aligned}
\label{eqSB9}
\end{equation}
These can be written in the unified form for arbitrary site $i$ reads
\begin{equation}
\begin{aligned}
    E_{i} = i g + (-1)^{i+1} \frac{\delta_{\rm{medium}}}{2}.
\end{aligned}
\label{eqSB10}
\end{equation}
where the subband offset $\delta_{\rm{medium}}$ is
\begin{equation}
    \begin{aligned}
    \delta_{\rm{medium}} & = 2\Delta - 2E_{n,-}^{(2)}  + g\\
    &=2\Delta + g - \frac{t^{2}\left[(B_{\mathrm{z},1}-g)(B_{\mathrm{z},1}-3g)+2t^{2}\right]}{2\Delta\left[(3g-B_{\mathrm{z},1})(g+B_{\mathrm{z},1})-4t^{2}\right]}.
    \end{aligned}
    \label{eqSB11}
\end{equation}

The nearest-neighbor energy gaps between adjacent dimers become:
\begin{equation}
    \begin{aligned}
    \Delta E_1^{\text{odd}\to\text{even}} 
    & = E_{2l+1} -E_{2l}  =g+ \delta_{\rm{medium}} \\
    \Delta E_1^{\text{even}\to\text{odd}} 
    &= E_{2l} -E_{2l-1}   = g- \delta_{\rm{medium}}  \\
    \end{aligned}
    \label{eqSB12}
\end{equation}
While the next-nearest-neighbor gap remains
\begin{equation}
    \begin{aligned}
    \Delta E_2=2g.
    \end{aligned}
    \label{eqSB13}
\end{equation}

In the strong-field limit ($B_{\mathrm{z},1} \gg g$), where $\Delta \approx (B_{\mathrm{z},1}-g)/2$, $\sin\theta \approx t/(B_{\mathrm{z},1}-g) \ll 1$, and $\cos\theta \approx 1$, this reduces to the result quoted in Eq.~\eqref{eq10} of the main text:
\begin{equation}
    \begin{aligned}
    \delta_{\mathrm{strong}} = B_{\mathrm{z},1} + \frac{2t^{2}}{g+B_{\mathrm{z},1}} + \mathcal{O} \left( \frac{t^{4}}{B_{\mathrm{z},1}^{3}} \right).
    \end{aligned}
    \label{eqSB14}
\end{equation}

\section{Perturbative calculation of $k_{\rm{eff}}$ in the weak-field limit}
\label{app-C}
\setcounter{equation}{0}
\renewcommand{\theequation}{D\arabic{equation}} 

Now we consider the regime where a purely quadratic magnetic gradient is present, i.e., $B_{z,2} \ll t, g$ and $B_{z,1} = 0$. The unperturbed Hamiltonian $\hat{H}_0$ is identical to that in Eq.~\eqref{eqSA2} with eigenenergies $E_n^{(0)} = ng$ and eigenstates given by the Wannier--Stark states $|\psi_n^{(0)}\rangle$, and the perturbation arising from the quadratic gradient reads
\begin{equation}
\hat{V}_{\mathrm{eff},2} = B_{z,2} \sum_{i=1}^{N} i \times (-1)^{i+1} \, |i\rangle \langle i| \,,
\label{eqSC1}
\end{equation}

In the basis $\{ \left| \psi_n^{(0)} \right\rangle \}$, the matrix elements of perturbation are 
\begin{equation}
    \begin{aligned}
    V_{n,m} 
    &\equiv \left\langle {\psi_n ^{(0)} } \left|   V_{\mathrm{eff},2}  \right| {\psi_m ^{(0)} } \right\rangle   \\
    &=  B_{z,2}  \sum_{i=1}^N  i \times (-1)^{i+1} \mathcal{J}_{i-n}   \left(\frac{2t}{g}\right) \mathcal{J}_{i-m}  \left(\frac{2t}{g}\right) ,
    \end{aligned}
    \label{eqSC2}
\end{equation}
In the thermodynamic limit $N \to \infty$ and let $k=i-n$, $\alpha=4t/g$, the summation becomes 
\begin{equation}
    \begin{aligned}
    V_{n,m} =  B_{z,2}  \sum_{k=-\infty}^{\infty}  (k+n) \times (-1)^{k+n+1} \mathcal{J}_{k}   \left(\frac{\alpha}{2}\right) \mathcal{J}_{k+n-m}  \left(\frac{\alpha}{2}\right),
    \end{aligned}
    \label{eqSC3}
\end{equation}
Using the identities for Bessel functions, 
\begin{equation}
    \begin{aligned}
    & \sum_{k=-\infty}^{\infty} (-1)^{k} \mathcal{J}_{k}^{2}(x) = \mathcal{J}_0(2x)  \\ 
    & \sum_{k=-\infty}^{\infty} k   (-1)^{k} \mathcal{J}_{k}^{2}(x) = 0 ,
    \end{aligned}
    \label{eqSC4}
\end{equation}
the expression simplifies to
\begin{equation}
    \begin{aligned}
    V_{n,m} = B_{z,2} \, (-1)^{n+1} \, (-1)^{m-n} \, \mathcal{J}_{m-n}(\alpha) \left(  \frac{m+n}{2} \right).
    \end{aligned}    
    \label{eqSC5}
\end{equation}

Then the first-, second-, and third-order corrections to the eigenenergies are
\begin{equation}
    \begin{aligned}
    E_n^{(1)} 
    = V_{n,n}
    = B_{z,2} \, n \, (-1)^{n+1} \mathcal{J}_0   \left( \frac{4t}{g} \right),
    \end{aligned} 
    \label{eqSC6}
\end{equation}
\begin{equation}
    \begin{aligned}
    E_n^{(2)} 
    &= \sum_{p \neq 0} \frac{|V_{n,n+p}|^{2}}{E_n^{(0)} - E_{n+p}^{(0)}} \\
    &= -\frac{B_{z,2}^{2}}{g} \sum_{p \neq 0} \frac{\mathcal{J}_{p}^{2}(\alpha)}{p} \left( n + \frac{p}{2} \right)^{2} \\
    &= -\frac{B_{z,2}^{2}}{g} \, \bigl[1 - \mathcal{J}_0^{2}(\alpha)\bigr] \, n ,
    \end{aligned} 
    \label{eqSC7}
\end{equation}
and
\begin{equation}
    \begin{aligned}
    E_n^{(3)} 
    &= \sideset{}{'}\sum_{p,q} \frac{V_{n,n+p} \, V_{n+p,n+q} \, V_{n+q,n}}{(-pg)(-qg)} \\ 
    & - V_{nn} \sideset{}{'}\sum_{p} \frac{|V_{n,n+p}|^{2}}{(-pg)^{2}} \\
    & =  \frac{B_{z,2}^{3}}{g^{2}} \, (-1)^{n+1} \Bigl[ \mathcal{C}_{3a}(\alpha) \, n^{3} + \mathcal{C}_{3b}(\alpha) \, n \Bigr],
    \end{aligned} 
    \label{eqSC8}
\end{equation}
where the dimensionless coefficients are given by
\begin{equation}
    \begin{aligned}
    \mathcal{C}_{3a} &=  \sideset{}{'}\sum_{p,q} \frac{(-1)^{p} \mathcal{J}_{p}(\alpha) \mathcal{J}_{q-p}(\alpha) \mathcal{J}_{q}(\alpha)}{pq}      \\
    &- \mathcal{J}_0(\alpha) \sideset{}{'}\sum_{p} \frac{\mathcal{J}_{p}^{2}(\alpha)}{p^{2}}  \,, \\
    \mathcal{C}_{3b} &=  \sideset{}{'}\sum_{p,q} \frac{(-1)^{p} \mathcal{J}_{p}(\alpha) \mathcal{J}_{q-p}(\alpha) \mathcal{J}_{q}(\alpha)}{4pq} \, \bigl( p^{2} + 3pq + q^{2} \bigr)  \\
    &- \mathcal{J}_0(\alpha) \frac{1}{4} \sideset{}{'}\sum_{p} \mathcal{J}_{p}^{2}(\alpha) \,.
    \end{aligned} 
    \label{eqSC9}
\end{equation}
Note that $\frac{1}{4}\sum_{p}' \mathcal{J}_{p}^{2}(\alpha)=\frac{1-\mathcal{J}_{0}^{2}(\alpha)}{4}$.

It is convenient to introduce the parity-independent slope renormalization arising from the second-order correction,
\begin{equation}
    \begin{aligned}
    k_c = \frac{B_{z,2}^{2}}{g} \, \bigl[1 - \mathcal{J}_0^{2}(\alpha)\bigr] \,,
    \end{aligned} 
    \label{eqSC10}
\end{equation}
so that $E_n^{(2)}=-k_c n$.

Collecting all contributions up to third order, the total eigenenergy reads
\begin{equation}
    \begin{aligned}
    E_n &= ng + (-1)^{n+1} n \left( k_{\mathrm{eff}}^{(1)}  + \frac{B_{z,2}^{3}}{g^{2}} \, \mathcal{C}_{\Sigma}  \right) \\
    &\quad - k_c \, n + \frac{B_{z,2}^{3}}{g^{2}} \, (-1)^{n+1} \, \mathcal{C}_{3a} \, n^{3} ,
    \end{aligned}
    \label{eqSC11}
\end{equation}
where $k_{\mathrm{eff}} $ and $k_c $ are given by Eqs.~(\ref{eqSC6}) and (\ref{eqSC10}), respectively, and $\mathcal{C}_{\Sigma} = \mathcal{C}_{3b} + \mathcal{J}_0(\alpha)[1-\mathcal{J}_0^{2}(\alpha)]/4$ collects the linear third-order contribution. 

The nearest-neighbor energy gaps between adjacent rungs of the Wannier-Stark ladder are
\begin{equation}
    \begin{aligned}
    \Delta E_{1}^{+}(n) &= E_{2n+1} - E_{2n} \\
    & = g + (4n+1) \, k_{\mathrm{eff}} - k_c + \mathcal{O}(B_{z,2}^{3}) \,,  \\
    \Delta E_{1}^{-}(n) &= E_{2n+2} - E_{2n+1} \\
    & = g - (4n+3) \, k_{\mathrm{eff}} - k_c + \mathcal{O}(B_{z,2}^{3}) \,. 
    \end{aligned}
    \label{eqSC12}
\end{equation}
The first-order term $k_{\mathrm{eff}}$ produces the scissor-like shear discussed in Sec.~\ref{section:III-B}, with slopes $g \pm k_{\mathrm{eff}}$ on the two subladders. 
The second-order correction $k_c$ introduces a uniform shift $-k_c$ to both gaps, destroying the parallelism of the subladder branches while preserving their average spacing at $g$. 
The third-order $n^3$ term generates an $n^2$ contribution to the gaps, which is essential for capturing the curvature of the spectral fanning at high excitation numbers ($n \gtrsim 10$) observed in exact diagonalization. 
This nonlinear $n$-dependence of the nearest-neighbor spacing directly manifests as the site-dependent beating patterns discussed in the main text, with modulation periods scaling as $T_{\mathrm{beat}} \propto 1/(2n+1)$.

\section{Renormalization of $\delta$ and $k_{\rm{eff}}$ in the weak-field limit}
\label{app-D}
\setcounter{equation}{0}
\renewcommand{\theequation}{E\arabic{equation}} 

We now consider the regime where both linear and quadratic magnetic gradients are present simultaneously,
i.e., $ B_{\mathrm{z,1}}, B_{\mathrm{z,2}} \ll t, g $ and the chain length $ N \to \infty$.
The unperturbed Hamiltonian $\hat{H}_0$ is identical to that in Appendix~\ref{app-A} and Appendix~\ref{app-C}, with eigenenergies $E_n^{(0)} = ng$ and eigenstates given by the Wannier--Stark states $|\psi_n^{(0)}\rangle$.
The perturbation is a superposition of the staggered potentials arising from both gradients:
\begin{equation}
    \begin{aligned}
    \hat{V}_{\mathrm{eff}} & = V_{\rm{eff,1}} + V_{\rm{eff,2}} \\
    &=  \sum_{i=1}^{N} (-1)^{i+1}  \left[ \frac{B_{\mathrm{z,1}}}{2} +B_{\mathrm{z,2}} \times i  \right]  | i \rangle \langle i | .
    \end{aligned}
    \label{eqSD1}
\end{equation}

In the basis $\{ \left| \psi_n^{(0)} \right\rangle \}$, adding Eqs.~\eqref{eqSA7} and \eqref{eqSC5}, the matrix elements of $\hat{V}_{\mathrm{eff}}$ are then
\begin{equation}
    \begin{aligned}
    V_{n,n+p} 
          & = \langle \psi_{n+p}^{(0)} | \hat{V}_{\rm eff} | \psi_n^{(0)} \rangle \\
          & = \langle \psi_{n+p}^{(0)} | V_{\rm{eff,1}} | \psi_n^{(0)} \rangle + \langle \psi_{n+p}^{(0)} | V_{\rm{eff,2}} | \psi_n^{(0)} \rangle \\
          & =  (-1)^{n+1}(-1)^{p}\mathcal{J}_{p}(\alpha)\left[\frac{B_{\mathrm{z,1}}}{2} + B_{\mathrm{z,2}}\left(n+\frac{p}{2}\right)\right],
    \end{aligned}
    \label{eqSD2}
\end{equation}
where $\alpha=4t/g$. 
This unified expression correctly reduces to the pure-linear result of Appendix~\ref{app-A} when $B_{\mathrm{z,2}}=0$, and to the pure-quadratic result of Appendix~\ref{app-C} when $B_{\mathrm{z,1}}=0$.
For brevity, the argument $\alpha=4t/g$ of the Bessel functions is omitted in the following: $\mathcal{J}_p\equiv\mathcal{J}_p(\alpha)$.

The first-order correction is simply the sum of the individual first-order corrections derived in Appendix~\ref{app-A} and Appendix~\ref{app-C}:
\begin{equation}
    \begin{aligned}
    V_{n,n} 
          & = \langle \psi_n^{(0)} | V_{\rm eff,1} | \psi_n^{(0)} \rangle
          + \langle \psi_n^{(0)} | V_{\rm eff,2} | \psi_n^{(0)} \rangle \\
          & = (-1)^{n+1}\mathcal{J}_0\left[\frac{B_{\rm z,1}}{2} + n B_{\rm z,2}\right] \\
          & = (-1)^{n+1}\left(\delta_{\rm weak} + n k_{\rm eff}\right).
    \end{aligned}
    \label{eqSD3}
\end{equation}

The second-order energy correction is
\begin{equation}
    \begin{aligned}
       E_n^{(2)} 
       & = \sum_{p \neq 0} \frac{|V_{n,n+p}|^2}{E_n^{(0)} - E_{n+p}^{(0)}} \\
       & = -\frac{1}{g}\sum_{p\neq 0}\frac{\mathcal{J}_p^2}{p}
            \left[\frac{B_{\mathrm{z,1}}}{2} + B_{\mathrm{z,2}}\left(n+\frac{p}{2}\right)\right]^2.
    \end{aligned}
    \label{eqSD4}
\end{equation}

Expanding the square and using the identities (valid for infinite sums over $p\neq 0$)
\begin{equation}
    \begin{aligned}
    &\sum_{p\neq 0}\frac{\mathcal{J}_p^2}{p}=0,  \\
    &\sum_{p\neq 0}\mathcal{J}_p^2=1-\mathcal{J}_0^2,  \\
    &\sum_{p\neq 0}p\,\mathcal{J}_p^2=0,
    \end{aligned}
    \label{eqSD5}
\end{equation}
only the cross-term proportional to $B_{\mathrm{z,2}}\,p$ inside the square contributes to a non-zero result. 

Then, expanding the square in Eq.~\eqref{eqSD4} and using the identities \eqref{eqSD5}, the second-order energy correction naturally decomposes into a pure-quadratic piece and a cross-gradient piece,
\begin{equation}
\begin{aligned}
E_n^{(2)} &= E_n^{(2,\,\text{pure})} + E_n^{(2,\,\text{cross})} \\
&= -n k_{\rm c} -\frac{1-\mathcal{J}_0^2(\alpha)}{2}\,\frac{B_{\rm z,1}B_{\rm z,2}}{g}.
\end{aligned}
\label{eqSD6}
\end{equation}
Here the parity-independent slope renormalization arising from the pure-quadratic contribution is
\begin{equation}
k_{\rm c} = \frac{B_{\rm z,2}^2}{g}\left[1-\mathcal{J}_0^2(\alpha)\right],
\label{eqSD7}
\end{equation}
so that $E_n^{(2,\,\text{pure})}=-nk_{\rm c}$ modifies the effective tilt equally for all states. The second term $E_n^{(2,\,\text{cross})}$ represents a global energy shift proportional to $B_{\rm z,1}B_{\rm z,2}$ that stems directly from the cross coupling between the two gradients.

To capture the remaining non-additive effects arising from the interference of virtual transitions induced by both gradients, we extend the perturbative calculation to the third order. 
The third-order energy correction is
\begin{equation}
    \begin{aligned}
    E_n^{(3)} 
    & = \sideset{}{'}\sum_{p,q}\frac{V_{n,n+p}V_{n+p,n+q}V_{n+q,n}}{(-pg)(-qg)}
          - V_{nn}\sideset{}{'}\sum_p\frac{|V_{n,n+p}|^2}{(-pg)^2} \\
    & = \frac{(-1)^{n+1}}{g^2} \sideset{}{'}\sum_{p,q} \frac{(-1)^p\mathcal{J}_p\mathcal{J}_{q-p}\mathcal{J}_q}{pq}\, \mathcal{F}_{n,n+p}\mathcal{F}_{n+p,n+q}\mathcal{F}_{n+q,n} \\
    &- \frac{(-1)^{n+1}}{g^2}\mathcal{J}_0\mathcal{F}_{nn}\sideset{}{'}\sum_p\frac{\mathcal{J}_p^2\,\mathcal{F}_{n,n+p}^2}{p^2}
    \end{aligned}
    \label{eqSD8}
\end{equation}
where the primed sums $\sum'$ exclude vanishing denominators.
And we have defined
\begin{equation}
    \begin{aligned}
    \mathcal{F}_{n,n+p} = \frac{B_{\mathrm{z,1}}}{2} + B_{\mathrm{z,2}}\left(n+\frac{p}{2}\right).
    \end{aligned}
    \label{eqSD9}
\end{equation}
Expanding the product of the three $\mathcal{F}$ factors in Eq.~\eqref{eqSD8} and collecting terms by powers of $B_{\mathrm{z,1}}$ and $B_{\mathrm{z,2}}$, the third-order correction naturally decomposes into pure-gradient and cross-gradient pieces,
\begin{equation}
E_n^{(3)} = E_n^{(3,\,\text{pure})} + E_n^{(3,\,\text{cross})}.
\label{eqSD10}
\end{equation}
The pure-gradient part $E_n^{(3,\,\text{pure})}$ reduces to the isolated results of Eq.~\eqref{eqSA10} and \eqref{eqSC8} when either $B_{\mathrm{z,2}}=0$ or $B_{\mathrm{z,1}}=0$, respectively. The cross-gradient part, which embodies the non-additive interference of virtual transitions induced by both gradients, reads
\begin{equation}
E_n^{(3,\,\text{cross})} = \frac{B_{\mathrm{z,1}}^2B_{\mathrm{z,2}}}{g^2}(-1)^{n+1}n\,C_4 + \frac{B_{\mathrm{z,1}}B_{\mathrm{z,2}}^2}{g^2}(-1)^{n+1}C_5,
\label{eqSD11}
\end{equation}
with
\begin{equation}
C_4 = \frac{3}{4}C(\alpha),
\label{eqSD12}
\end{equation}
and
\begin{equation}
\begin{aligned}
C_5 &= C_5^{(0)}(\alpha) + C_5^{(2)}(\alpha)\,n^2 \\
&= \frac{3}{8}\sideset{}{'}\sum_{p,q}\frac{(-1)^p\mathcal{J}_p(\alpha)\mathcal{J}_{q-p}(\alpha)\mathcal{J}_q(\alpha)}{pq}(p^2+pq+q^2) \\
&\quad -\frac{3\mathcal{J}_0(\alpha)}{8}\sideset{}{'}\sum_p\frac{\mathcal{J}_p^2(\alpha)}{p^2} + \frac{3C(\alpha)}{2}n^2,
\end{aligned}
\label{eqSD13}
\end{equation}
where $C(\alpha)$ is exactly the same coefficient appearing in the pure-linear third-order result [Eq.~\eqref{eqSA11}].

Together with the second-order cross-term present in Eq.~\eqref{eqSD6}, these contributions quantify the failure of simple superposition: the $B_{\mathrm{z,1}}^2B_{\mathrm{z,2}}$ term renormalizes the subladder slopes differentially through $C_4$, while the $B_{\mathrm{z,1}}B_{\mathrm{z,2}}^2$ term introduces a progressive $n^2$ warping through $C_5^{(2)}$ that drives high-lying states away from linear extrapolation.

Collecting all contributions up to third order, the total coupled energy takes the compact form Eq.~\eqref{eq16b} in the main text:
\begin{equation}
    \begin{aligned}
    E_n^{(\text{cou})} & =  E_n^{(2,\,\text{cross})} +E_n^{(3,\,\text{cross})} \\
    & = C_3\frac{B_{\mathrm{z,1}}B_{\mathrm{z,2}}}{g} + C_4\frac{B_{\mathrm{z,1}}^2B_{\mathrm{z,2}}}{g^2}n(-1)^{n+1} + C_5\frac{B_{\mathrm{z,1}}B_{\mathrm{z,2}}^2}{g^2}(-1)^{n+1},
    \end{aligned}
\label{eqSD14}
\end{equation}
where $C_3=-\frac{1}{2}[1-\mathcal{J}_0^2(\alpha)]$ originates from the second-order cross-term in Eq.~\eqref{eqSD6}, while $C_4$ and $C_5$ are given by Eqs.~\eqref{eqSD12} and \eqref{eqSD13}, respectively. Note that although $C_3$ and $C_4$ are genuine constants, $C_5$ acquires a quadratic $n$-dependence through $C_5^{(2)}$, reflecting the asymptotic bending of the Wannier--Stark ladder at large excitation numbers.

\section{Leakage-induced renormalization of the Wannier--Stark ladder and Bloch-oscillation peaks}
\label{app-F}
\setcounter{equation}{0}
\renewcommand{\theequation}{F\arabic{equation}}
In this appendix, we translate the leakage-induced renormalization obtained in Appendix~\ref{app-E} into explicit corrections to the WS ladder and to the BO peak positions analyzed in Appendix~\ref{app-D}.
The leakage correction given by Eq.~\eqref{eqSE7} can be incorporated into the on-site potential of the effective single-particle model, Eq.~\eqref{eq3}, which takes the form
\begin{equation}
    \widetilde V(i)= g i + \frac{ (-1)^{i+1} B_{z,1}}{2} + (-1)^{i+1}B_{z,2} \times i + \delta V_{\rm leak}(i),
\label{eqSF1}
\end{equation}
with
\begin{equation}
    \delta V_{\rm leak}(i)= (-1)^{i+1}\kappa_{\rm eff}B_{z,2} + (A_{\rm cross}+B_{\rm cross}\,i)B_{z,1}B_{z,2}.
\label{eqSF2}
\end{equation}
The coefficients $\kappa_{\rm eff}$ and $B_{\rm cross}$ are defined in Eq.~(\ref{eqSE8}) and Eq.~(\ref{eqSE9}); the constant $A_{\rm cross}$ can be absorbed into the energy offset and does not affect level spacings.
\subsection{Parameter renormalization by virtual hole--doublon pairs}
\label{app-F1}
Rearranging Eq.~\eqref{eqSF1} gives
\begin{equation}
    \widetilde V(i)= (g+\gamma_{\rm L})i + (-1)^{i+1} \left[\frac{B_{z,1}+\eta_{\rm L}}{2}+B_{z,2}i\right] + {\rm const},
    \label{eqSF3}
\end{equation}
with the two effective parameters
\begin{equation}
    \begin{aligned}
    & \eta_{\rm L}\equiv 2\kappa_{\rm eff}B_{z,2} ,\\
    & \gamma_{\rm L}\equiv B_{\rm cross}B_{z,1}B_{z,2}.
    \end{aligned}    
\label{eqSF4}
\end{equation}
Thus, to second order in the hopping amplitude, the effect of virtual hole--doublon excitations outside the truncated Hilbert space can be absorbed into the effective single-defect model as
\begin{equation}
    \begin{aligned}
    & g\;\to\; g+\gamma_{\rm L},\\
    & B_{z,1}\;\to\; B_{z,1}+\eta_{\rm L},
    \end{aligned}  
\label{eqSF5}
\end{equation}
while the quadratic gradient $B_{z,2}$ remains unchanged.
All analytic expressions for the WS spectrum and the BO peaks obtained in Appendix~\ref{app-D} can therefore be carried over by simply applying the replacements Eq.~(\ref{eqSF5}).

\subsection{Linear gradient only ($B_{z,2}=0$)}
\label{app-F2}
When the quadratic gradient is absent, $\eta_{\rm L}=\gamma_{\rm L}=0$ and the parameters are unrenormalized.
The Bloch peaks are still given by Eq.~(\ref{eqSA14}),
\begin{equation}
    \omega_{1,\pm}=g\pm\delta_{\rm weak},\qquad \omega_2=2g,
    \label{eqSF6}
\end{equation}
and leakage induces no frequency shift.

\subsection{Quadratic gradient only ($B_{z,1}=0$)}
\label{app-F3}
For a purely quadratic magnetic field, the leakage potential reduces to an effective staggered field of amplitude $B_{z,1}^{\rm (ind)}=\eta_{\rm L}$.
Inserting this induced field into the expression for the subladder offset $\delta_{\rm weak}$ [Eq.~(\ref{eqSA14})] and retaining only the leading linear term gives
\begin{equation}
    \begin{aligned}
    \delta_{\rm L}^{(0)} & \equiv\delta_{\rm weak}(\eta_{\rm L})\simeq \eta_{\rm L}\,\mathcal J_0 \left(\frac{4t}{g}\right) \\
    & = 2\kappa_{\rm eff}B_{z,2}\,\mathcal J_0 \left(\frac{4t}{g}\right).
    \end{aligned}    
\label{eqSF7}
\end{equation}
Combining this with the pure quadratic-gradient results of Appendix~\ref{app-C}, the nearest-neighbor gaps become
\begin{equation}
\begin{aligned}
\omega_{1,2\ell} &= g - k_c + (4\ell+1)k_{\rm eff} + \delta_{\rm L}^{(0)},\\
\omega_{1,2\ell+1} &= g - k_c - (4\ell+3)k_{\rm eff} - \delta_{\rm L}^{(0)},
\end{aligned}
\label{eqSF8}
\end{equation}
where $k_{\rm eff}=B_{z,2}\mathcal J_0(4t/g)$ and $k_c=(B_{z,2}^2/g)[1-\mathcal J_0^2(4t/g)]$.
The two fundamental branches are therefore shifted symmetrically by $\pm\delta_{\rm L}^{(0)}$.
The next-nearest-neighbor gaps read
\begin{equation}
    \omega_{2,\pm}=2(g-k_c\pm k_{\rm eff}),
    \label{eqSF9}
\end{equation}
and are unaffected by leakage because the alternating offset cancels within the same subladder.

\subsection{Combined gradients ($B_{z,1}\neq0,\;B_{z,2}\neq0$)}
\label{app-F4}
In the presence of both gradients, the replacements Eq.~(\ref{eqSF5}) must be applied to the coupled-gradient energy $E_n^{\rm (coupled)}$ derived in Appendix~\ref{app-D}:
\begin{equation}
    E_n^{\rm (full)} = E_n^{\rm (coupled)}\big|_{g\to g+\gamma_{\rm L},\, B_{z,1}\to B_{z,1}+\eta_{\rm L}}.
\label{eqSF10}
\end{equation}
Since $\gamma_{\rm L}$ and $\eta_{\rm L}$ are already second-order leakage parameters, the replacements in Eq.~(\ref{eqSF10}) are understood only to linear order in $\gamma_{\rm L}$ and $\eta_{\rm L}$.
Expanding to first order in the leakage parameters yields the shifts of the nearest-neighbor gaps,
\begin{equation}
    \Delta\omega_{1,n}^{\rm (leak)} =
    \begin{cases}
        \gamma_{\rm L} + \delta_{\rm L}, & n\ \text{even},\\[2pt]
        \gamma_{\rm L} - \delta_{\rm L}, & n\ \text{odd},
    \end{cases}
    \label{eqSF11}
\end{equation}
where
\begin{equation}
    \begin{aligned}
    \delta_{\rm L} & \equiv\delta_{\rm weak}(B_{z,1}+\eta_{\rm L})-\delta_{\rm weak}(B_{z,1})  \\
    & \simeq 2\kappa_{\rm eff}B_{z,2} \left[ \mathcal J_0\left(\frac{4t}{g}\right) + \frac{3B_{z,1}^2}{4g^2} C\left(\frac{4t}{g}\right) \right].
    \end{aligned}
    \label{eqSF12}
\end{equation}
The next-nearest-neighbor gaps are shifted by a uniform amount,
\begin{equation}
\Delta\omega_2^{\rm (leak)} = 2\gamma_{\rm L} = 2B_{\rm cross}B_{z,1}B_{z,2},
\label{eqSF13}
\end{equation}
as the alternating terms cancel for intrasubladder transitions.

\subsection{Magnitude of the corrections}
\label{app-F5}
To combine the leakage corrections derived above with the intrinsic spectral formulas of Appendix~\ref{app-D}, it is convenient to introduce three effective parameters
\begin{equation}
    \begin{aligned}
    & G = g+\gamma_{\rm L}-k_c,\\
    & D = \delta_{\rm weak}+\delta_{\rm L},\\
    & K = k_{\rm eff}+C_4\frac{B_{z,1}^2B_{z,2}}{g^2},
    \end{aligned}
\label{eqSF14}
\end{equation}
where $C_4$ is the cross-gradient slope renormalization derived in Appendix~\ref{app-D}.
We note that $D$ collects all parity-dependent, $n$-independent offsets; the remaining third-order constant terms from Appendix~\ref{app-D} can be absorbed into $D$ without affecting the order-of-magnitude estimates below.
Neglecting the high-order warping terms, the fundamental and first-harmonic Bloch peaks take the compact form
\begin{equation}
    \begin{aligned}
    \omega_{1,n} &= G + D + K(2n+1)\quad (n\ \text{even}),\\
    \omega_{1,n} &= G - D - K(2n+1)\quad (n\ \text{odd}),
    \end{aligned}
    \label{eqSF15}
\end{equation}
and 
\begin{equation}
    \begin{aligned}
    \omega_{2,n} &= 2(G - K)\quad (n\ \text{even}),\\
    \omega_{2,n} &= 2(G + K)\quad (n\ \text{odd}).
    \end{aligned}
    \label{eqSF16}
\end{equation}
In these expressions, leakage enters exclusively through the two small quantities $\gamma_{\rm L}$ and $\delta_{\rm L}$, which shift $G$ and $D$ but do not affect $K$.  
Consequently, leakage can at most produce a rigid displacement of the entire spectrum (via $\gamma_{\rm L}$) and a symmetric shift of the fundamental doublet branches (via $\delta_{\rm L}$), while the leakage-induced contribution to the first-harmonic doublet splitting vanishes.  
The intrinsic cross-gradient term contained in $K$ can nevertheless modify this splitting.

We now estimate the magnitude of these terms.  
Using Eq.~(\ref{eqSE8}) with $w_X\simeq 2/X^2$ (valid for $g\ll\Delta_{\rm{leak}},\Delta_{\rm{leak}}'$), 
the observable leakage-induced peak shifts scale as
\begin{equation}
    \begin{aligned}
    &|\delta_{\rm L}|\simeq 4\,\mathcal{J}_0 \left(\frac{4t}{g}\right) 
    t^{2}B_{z,2}\left(\frac{3}{\Delta_{\rm{leak}}^{2}}-\frac{2}{\Delta_{\rm{leak}}'^{2}}\right),\\
    &|\gamma_{\rm L}|\simeq 4t^{2}B_{z,1}B_{z,2}
    \left(\frac{2}{\Delta_{\rm{leak}}'^{3}}-\frac{1}{\Delta_{\rm{leak}}^{3}}\right),
    \end{aligned}
    \label{eqSF17}
\end{equation}
where $\Delta_{\rm{leak}}$ and $\Delta_{\rm{leak}}'$ are the two characteristic leakage gaps defined in Eq.~(\ref{eqSE5}).
Note that the induced staggered field $\eta_{\rm L}$ enters the peak positions only through $\delta_{\rm weak}$ [Eq.~(\ref{eqSA14})], whose linear coefficient $\mathcal{J}_0(4t/g)$ depends only on $t/g$ and therefore does not affect the scaling with $J_z/t$.
For the main-text choice $U=2J_z$, i.e., $\Delta_{\rm{leak}}=5J_z/2$ and $\Delta_{\rm{leak}}'=11J_z/4$, the staggered shift scales as $|\delta_{\rm L}|\propto t^{2}B_{z,2}/J_z^{2}$ while the tilt shift scales as $|\gamma_{\rm L}|\propto t^{2}B_{z,1}B_{z,2}/J_z^{3}$.
For the representative strong-coupling parameters used in the main text ($U=200g$, $J_z=100g$, $t=0.5g$, $B_{z,1}=0.2g$, $B_{z,2}=10^{-3}g$) one finds
\begin{equation}
    \begin{aligned}
    & |\delta_{\rm L}^{(0)}|\sim 4.8 \times10^{-9}g,\\
    & |\delta_{\rm L}|\sim 5.5 \times10^{-9}g,\\
    & |\gamma_{\rm L}|\sim 6.4 \times10^{-12}g.
    \end{aligned}
    \label{eqSF18}
\end{equation}
These values are three to four orders of magnitude smaller than the nonadditive cross-coupling features ($\sim10^{-4}g$--$10^{-5}g$) analyzed in Sec.~\ref{section:III-C}.

Spinon excitations, which require a second hop from the HD intermediate state and are reached only at fourth order in $t$, contribute at order $t^4/\Delta_{\rm{leak}}^3$.  
They are therefore suppressed by an additional factor $t^2/\Delta_{\rm{leak}}^2\sim10^{-5}$ relative to the HD/DH corrections given above, and are negligible under the strong-coupling conditions of interest.

From Eq.~\eqref{eqSF7} and Eq.~\eqref{eqSF12}, the staggered leakage shift grows as $|\delta_{\rm L}|\propto t^2B_{z,2}/\Delta_{\rm{leak}}^2$ when $J_z/t$ is reduced, whereas the tilt shift grows only as $|\gamma_{\rm L}|\propto t^2B_{z,1}B_{z,2}/\Delta_{\rm{leak}}^3$; the staggered shift therefore sets the binding constraint.
Using $\Delta_{\rm{leak}}=5J_z/2$ and $\Delta_{\rm{leak}}'=11J_z/4$ for $U=2J_z$ in Eq.~(\ref{eqSE8}), one finds $|\delta_{\rm L}^{(0)} | \simeq0.86\,\mathcal{J}_0(4t/g)\,t^{2}B_{z,2}/J_z^{2}$.
For $t=0.5g$ and $B_{z,2}=10^{-3}g$, keeping $|\delta_{\rm L}^{(0)}|$ below the smallest intrinsic cross-coupling feature ($\sim10^{-7}g$) requires $J_z\gtrsim22g$, while keeping it one order of magnitude below requires $J_z\gtrsim70g$ [cf.~Fig.~\ref{figE2}(b)].
The main-text parameters ($J_z=100g$, $U=2J_z$) satisfy even the latter, stricter criterion: the leakage shifts $\sim5\times10^{-9}g$ lie more than one order of magnitude below the smallest intrinsic features ($\sim10^{-7}g$) and three to four orders below the dominant cross-coupling features ($10^{-5}g$--$10^{-4}g$).

The rigid baseline shift of the fundamental doublet and the asymmetric broadening of the first harmonic reported in the main text are therefore intrinsic consequences of the coupled magnetic gradients, not artifacts originating from virtual hole--doublon excitations outside the truncated single-defect Hilbert space.

\end{appendix}

\bibliography{refs}

@book{Giamarchi2003Book,
    author = {Giamarchi, Thierry},
    title = {Quantum Physics in One Dimension},
    publisher = {Oxford University Press},
    year = {2003},
    month = {12},
    isbn = {9780198525004},
    doi = {10.1093/acprof:oso/9780198525004.001.0001},
    url = {https://doi.org/10.1093/acprof:oso/9780198525004.001.0001},
}

@article{Maksimov2015PRA,
  title = {Wannier-Stark states in double-periodic lattices. I. One-dimensional lattices},
  author = {Maksimov, Dmitrii N. and Bulgakov, Evgeny N. and Kolovsky, Andrey R.},
  journal = {Phys. Rev. A},
  volume = {91},
  issue = {5},
  pages = {053631},
  numpages = {8},
  year = {2015},
  month = {May},
  publisher = {American Physical Society},
  doi = {10.1103/PhysRevA.91.053631},
  url = {https://link.aps.org/doi/10.1103/PhysRevA.91.053631}
}

@article{Dreisow2009PRL,
  title = {Bloch-Zener Oscillations in Binary Superlattices},
  author = {Dreisow, F. and Szameit, A. and Heinrich, M. and Pertsch, T. and Nolte, S. and T\"unnermann, A. and Longhi, S.},
  journal = {Phys. Rev. Lett.},
  volume = {102},
  issue = {7},
  pages = {076802},
  numpages = {4},
  year = {2009},
  month = {Feb},
  publisher = {American Physical Society},
  doi = {10.1103/PhysRevLett.102.076802},
  url = {https://link.aps.org/doi/10.1103/PhysRevLett.102.076802}
}

@article{Montorsi2020PRA,
  title = {Homogeneous and domain-wall topological Haldane conductors with dressed Rydberg atoms},
  author = {Montorsi, Arianna and Fazzini, Serena and Barbiero, Luca},
  journal = {Phys. Rev. A},
  volume = {101},
  issue = {4},
  pages = {043618},
  numpages = {7},
  year = {2020},
  month = {Apr},
  publisher = {American Physical Society},
  doi = {10.1103/PhysRevA.101.043618},
  url = {https://link.aps.org/doi/10.1103/PhysRevA.101.043618}
}

@article{PolkovnikovRMP2011,
  title = {Colloquium: Nonequilibrium dynamics of closed interacting quantum systems},
  author = {Polkovnikov, Anatoli and Sengupta, Krishnendu and Silva, Alessandro and Vengalattore, Mukund},
  journal = {Rev. Mod. Phys.},
  volume = {83},
  issue = {3},
  pages = {863--883},
  numpages = {0},
  year = {2011},
  month = {Aug},
  publisher = {American Physical Society},
  doi = {10.1103/RevModPhys.83.863},
  url = {https://link.aps.org/doi/10.1103/RevModPhys.83.863}
}

@article{AbaninRMP2019,
  title = {Colloquium: Many-body localization, thermalization, and entanglement},
  author = {Abanin, Dmitry A. and Altman, Ehud and Bloch, Immanuel and Serbyn, Maksym},
  journal = {Rev. Mod. Phys.},
  volume = {91},
  issue = {2},
  pages = {021001},
  numpages = {26},
  year = {2019},
  month = {May},
  publisher = {American Physical Society},
  doi = {10.1103/RevModPhys.91.021001},
  url = {https://link.aps.org/doi/10.1103/RevModPhys.91.021001}
}

@Article{Bloch1929,
author={Bloch, Felix},
title={{\"U}ber die Quantenmechanik der Elektronen in Kristallgittern},
journal={Zeitschrift f{\"u}r Physik},
year={1929},
month={Jul},
day={01},
volume={52},
number={7},
pages={555-600},
issn={0044-3328},
doi={10.1007/BF01339455},
url={https://doi.org/10.1007/BF01339455}
}

@article{Zener1934,
    author = {Zener, Clarence},
    title = {A theory of the electrical breakdown of solid dielectrics},
    journal = {Proceedings of the Royal Society of London. Series A, Containing Papers of a Mathematical and Physical Character},
    volume = {145},
    number = {855},
    pages = {523-529},
    year = {1934},
    month = {07},
    issn = {0950-1207},
    doi = {10.1098/rspa.1934.0116},
    url = {https://doi.org/10.1098/rspa.1934.0116},
}

@article{Houston1940PR,
  title = {Acceleration of Electrons in a Crystal Lattice},
  author = {Houston, W. V.},
  journal = {Phys. Rev.},
  volume = {57},
  issue = {3},
  pages = {184--186},
  numpages = {0},
  year = {1940},
  month = {Feb},
  publisher = {American Physical Society},
  doi = {10.1103/PhysRev.57.184},
  url = {https://link.aps.org/doi/10.1103/PhysRev.57.184}
}

@article{Wannier1960PR,
  title = {Wave Functions and Effective Hamiltonian for Bloch Electrons in an Electric Field},
  author = {Wannier, Gregory H.},
  journal = {Phys. Rev.},
  volume = {117},
  issue = {2},
  pages = {432--439},
  numpages = {0},
  year = {1960},
  month = {Jan},
  publisher = {American Physical Society},
  doi = {10.1103/PhysRev.117.432},
  url = {https://link.aps.org/doi/10.1103/PhysRev.117.432}
}

@article{Ben1996PRL,
  title = {Bloch Oscillations of Atoms in an Optical Potential},
  author = {Ben Dahan, Maxime and Peik, Ekkehard and Reichel, Jakob and Castin, Yvan and Salomon, Christophe},
  journal = {Phys. Rev. Lett.},
  volume = {76},
  issue = {24},
  pages = {4508--4511},
  numpages = {0},
  year = {1996},
  month = {Jun},
  publisher = {American Physical Society},
  doi = {10.1103/PhysRevLett.76.4508},
  url = {https://link.aps.org/doi/10.1103/PhysRevLett.76.4508}
}

@article{Wilkinson1996PRL,
  title = {Observation of Atomic Wannier-Stark Ladders in an Accelerating Optical Potential},
  author = {Wilkinson, S. R. and Bharucha, C. F. and Madison, K. W. and Niu, Qian and Raizen, M. G.},
  journal = {Phys. Rev. Lett.},
  volume = {76},
  issue = {24},
  pages = {4512--4515},
  numpages = {0},
  year = {1996},
  month = {Jun},
  publisher = {American Physical Society},
  doi = {10.1103/PhysRevLett.76.4512},
  url = {https://link.aps.org/doi/10.1103/PhysRevLett.76.4512}
}

@article{Feldmann1992PRB,
  title = {Optical investigation of Bloch oscillations in a semiconductor superlattice},
  author = {Feldmann, J. and Leo, K. and Shah, J. and Miller, D. A. B. and Cunningham, J. E. and Meier, T. and von Plessen, G. and Schulze, A. and Thomas, P. and Schmitt-Rink, S.},
  journal = {Phys. Rev. B},
  volume = {46},
  issue = {11},
  pages = {7252--7255},
  numpages = {0},
  year = {1992},
  month = {Sep},
  publisher = {American Physical Society},
  doi = {10.1103/PhysRevB.46.7252},
  url = {https://link.aps.org/doi/10.1103/PhysRevB.46.7252}
}

@ARTICLE{Esaki1970IBM,
  author={Esaki, L. and Tsu, R.},
  journal={IBM Journal of Research and Development}, 
  title={Superlattice and Negative Differential Conductivity in Semiconductors}, 
  year={1970},
  volume={14},
  number={1},
  pages={61-65},
  doi={10.1147/rd.141.0061}}

@article{Trompeter2006PRL,
  title = {Bloch Oscillations and Zener Tunneling in Two-Dimensional Photonic Lattices},
  author = {Trompeter, Henrike and Krolikowski, Wieslaw and Neshev, Dragomir N. and Desyatnikov, Anton S. and Sukhorukov, Andrey A. and Kivshar, Yuri S. and Pertsch, Thomas and Peschel, Ulf and Lederer, Falk},
  journal = {Phys. Rev. Lett.},
  volume = {96},
  issue = {5},
  pages = {053903},
  numpages = {4},
  year = {2006},
  month = {Feb},
  publisher = {American Physical Society},
  doi = {10.1103/PhysRevLett.96.053903},
  url = {https://link.aps.org/doi/10.1103/PhysRevLett.96.053903}
}

@article{Lin2007apl,
    author = {Lin, Weihua and Zhou, Xiang and Wang, Guo Ping and Chan, C. T.},
    title = {Spatial Bloch oscillations of plasmons in nanoscale metal waveguide arrays},
    journal = {Applied Physics Letters},
    volume = {91},
    number = {24},
    pages = {243113},
    year = {2007},
    month = {12},
    issn = {0003-6951},
    doi = {10.1063/1.2824394},
    url = {https://doi.org/10.1063/1.2824394},
}

@article{Moleron2019PRB,
  title = {Discrete propagation of trapped modes in acoustic waveguide arrays},
  author = {Moler\'on, Miguel and Faure, C\'edric and F\'elix, Simon and Pagneux, Vincent and Richoux, Olivier},
  journal = {Phys. Rev. B},
  volume = {99},
  issue = {20},
  pages = {201404},
  numpages = {6},
  year = {2019},
  month = {May},
  publisher = {American Physical Society},
  doi = {10.1103/PhysRevB.99.201404},
  url = {https://link.aps.org/doi/10.1103/PhysRevB.99.201404}
}

@article{Liu2021APE,
doi = {10.35848/1882-0786/ac0481},
url = {https://doi.org/10.35848/1882-0786/ac0481},
year = {2021},
month = {jun},
publisher = {IOP Publishing},
volume = {14},
number = {6},
pages = {064501},
author = {Liu, Yun-Kai and Wu, Hong-Wei and Hu, Ping and Sheng, Zong-Qiang},
title = {Spatial Bloch oscillations in acoustic waveguide arrays},
journal = {Applied Physics Express}
}

@article{Efremidis2002PRE,
  title = {Discrete solitons in nonlinear zigzag optical waveguide arrays with tailored diffraction properties},
  author = {Efremidis, Nikos K. and Christodoulides, Demetrios N.},
  journal = {Phys. Rev. E},
  volume = {65},
  issue = {5},
  pages = {056607},
  numpages = {11},
  year = {2002},
  month = {May},
  publisher = {American Physical Society},
  doi = {10.1103/PhysRevE.65.056607},
  url = {https://link.aps.org/doi/10.1103/PhysRevE.65.056607}
}

@article{TANG20173701PLA,
title = {Anharmonic Bloch oscillations in a helicoidal lattice},
journal = {Physics Letters A},
volume = {381},
number = {43},
pages = {3701-3705},
year = {2017},
issn = {0375-9601},
doi = {https://doi.org/10.1016/j.physleta.2017.09.036},
url = {https://www.sciencedirect.com/science/article/pii/S0375960117309015},
author = {Bing Tang and Xia Chang and Ke Deng}
}

@article{Haller2010PRL,
  title = {Inducing Transport in a Dissipation-Free Lattice with Super Bloch Oscillations},
  author = {Haller, Elmar and Hart, Russell and Mark, Manfred J. and Danzl, Johann G. and Reichs\"ollner, Lukas and N\"agerl, Hanns-Christoph},
  journal = {Phys. Rev. Lett.},
  volume = {104},
  issue = {20},
  pages = {200403},
  numpages = {4},
  year = {2010},
  month = {May},
  publisher = {American Physical Society},
  doi = {10.1103/PhysRevLett.104.200403},
  url = {https://link.aps.org/doi/10.1103/PhysRevLett.104.200403}
}

@article{Kudo2011PRA,
  title = {Theoretical analysis of super--Bloch oscillations},
  author = {Kudo, K. and Monteiro, T. S.},
  journal = {Phys. Rev. A},
  volume = {83},
  issue = {5},
  pages = {053627},
  numpages = {6},
  year = {2011},
  month = {May},
  publisher = {American Physical Society},
  doi = {10.1103/PhysRevA.83.053627},
  url = {https://link.aps.org/doi/10.1103/PhysRevA.83.053627}
}

@article{Zhu2019PRA,
  title = {Generalized Bloch oscillations of ultracold lattice atoms subject to higher-order gradients},
  author = {Zhu, Qian-Ru and Chen, Shou-Long and Li, Shao-Jun and Fang, Xue-Ting and Cao, Lushuai and Hu, Zhong-Kun},
  journal = {Phys. Rev. A},
  volume = {100},
  issue = {5},
  pages = {053603},
  numpages = {7},
  year = {2019},
  month = {Nov},
  publisher = {American Physical Society},
  doi = {10.1103/PhysRevA.100.053603},
  url = {https://link.aps.org/doi/10.1103/PhysRevA.100.053603}
}

@article{Khomeriki2010PRA,
  title = {Interaction-induced fractional Bloch and tunneling oscillations},
  author = {Khomeriki, Ramaz and Krimer, Dmitry O. and Haque, Masudul and Flach, Sergej},
  journal = {Phys. Rev. A},
  volume = {81},
  issue = {6},
  pages = {065601},
  numpages = {4},
  year = {2010},
  month = {Jun},
  publisher = {American Physical Society},
  doi = {10.1103/PhysRevA.81.065601},
  url = {https://link.aps.org/doi/10.1103/PhysRevA.81.065601}
}

@article{Cai2021PRL,
  title = {Multiparticle Quantum Walks and Fisher Information in One-Dimensional Lattices},
  author = {Cai, Xiaoming and Yang, Hongting and Shi, Hai-Long and Lee, Chaohong and Andrei, Natan and Guan, Xi-Wen},
  journal = {Phys. Rev. Lett.},
  volume = {127},
  issue = {10},
  pages = {100406},
  numpages = {7},
  year = {2021},
  month = {Sep},
  publisher = {American Physical Society},
  doi = {10.1103/PhysRevLett.127.100406},
  url = {https://link.aps.org/doi/10.1103/PhysRevLett.127.100406}
}

@article{Kyriakidis1998PRB,
  title = {Bloch oscillations of magnetic solitons in anisotropic spin-$\frac{1}{2}$ chains},
  author = {Kyriakidis, Jordan and Loss, Daniel},
  journal = {Phys. Rev. B},
  volume = {58},
  issue = {9},
  pages = {5568--5583},
  numpages = {0},
  year = {1998},
  month = {Sep},
  publisher = {American Physical Society},
  doi = {10.1103/PhysRevB.58.5568},
  url = {https://link.aps.org/doi/10.1103/PhysRevB.58.5568}
}

@article{Liu2019PRA,
  title = {Bloch oscillations of multimagnon excitations in a Heisenberg $XXZ$ chain},
  author = {Liu, Wenjie and Ke, Yongguan and Zhang, Li and Lee, Chaohong},
  journal = {Phys. Rev. A},
  volume = {99},
  issue = {6},
  pages = {063614},
  numpages = {10},
  year = {2019},
  month = {Jun},
  publisher = {American Physical Society},
  doi = {10.1103/PhysRevA.99.063614},
  url = {https://link.aps.org/doi/10.1103/PhysRevA.99.063614}
}

@Article{Hansen2022NC,
author={Hansen, Ursula B.
and Sylju{\aa}sen, Olav F.
and Jensen, Jens
and Sch{\"a}ffer, Turi K.
and Andersen, Christopher R.
and Boehm, Martin
and Rodriguez-Rivera, Jose A.
and Christensen, Niels B.
and Lefmann, Kim},
title={Magnetic Bloch oscillations and domain wall dynamics in a near-Ising ferromagnetic chain},
journal={Nature Communications},
year={2022},
month={May},
day={10},
volume={13},
number={1},
pages={2547},
issn={2041-1723},
doi={10.1038/s41467-022-29854-9},
url={https://doi.org/10.1038/s41467-022-29854-9}
}

@article{Kolezhuk2024PRB,
  title = {Current-controlled chirality dynamics in a mesoscopic magnetic domain wall},
  author = {Kolezhuk, Oleksiy and Teslia, Roman and Buryak, Ihor and Gomonay, Olena},
  journal = {Phys. Rev. B},
  volume = {109},
  issue = {13},
  pages = {134418},
  numpages = {10},
  year = {2024},
  month = {Apr},
  publisher = {American Physical Society},
  doi = {10.1103/PhysRevB.109.134418},
  url = {https://link.aps.org/doi/10.1103/PhysRevB.109.134418}
}

@Article{Xu2025NC,
author={Xu, Meng
and Hua, Chensong
and Chen, Yan
and Yu, Weichao},
title={Frequency modulation on magnons in synthetic dimensions},
journal={Nature Communications},
year={2025},
month={Apr},
day={09},
volume={16},
number={1},
pages={3356},
issn={2041-1723},
doi={10.1038/s41467-025-58582-z},
url={https://doi.org/10.1038/s41467-025-58582-z}
}

@article{Kosevich2013JPCM,
doi = {10.1088/0953-8984/25/24/246002},
url = {https://doi.org/10.1088/0953-8984/25/24/246002},
year = {2013},
month = {may},
publisher = {IOP Publishing},
volume = {25},
number = {24},
pages = {246002},
author = {Kosevich, Yuriy A and Gann, Vladimir V},
title = {Magnon localization and Bloch oscillations in finite Heisenberg spin chains in an inhomogeneous magnetic field},
journal = {Journal of Physics: Condensed Matter}
}

@article{Kohn1959PR,
  title = {Analytic Properties of Bloch Waves and Wannier Functions},
  author = {Kohn, W.},
  journal = {Phys. Rev.},
  volume = {115},
  issue = {4},
  pages = {809--821},
  numpages = {0},
  year = {1959},
  month = {Aug},
  publisher = {American Physical Society},
  doi = {10.1103/PhysRev.115.809},
  url = {https://link.aps.org/doi/10.1103/PhysRev.115.809}
}

@article{Zhang2026PRL,
  title = {Quantum Many-Body Dynamics for Fermionic $t\text{\ensuremath{-}}J$ Model Simulated with Atom Arrays},
  author = {Zhang, Ye-Bing and Zhou, Xin-Chi and Wang, Bao-Zong and Liu, Xiong-Jun},
  journal = {Phys. Rev. Lett.},
  volume = {136},
  issue = {3},
  pages = {033402},
  numpages = {9},
  year = {2026},
  month = {Jan},
  publisher = {American Physical Society},
  doi = {10.1103/px63-dtc9},
  url = {https://link.aps.org/doi/10.1103/px63-dtc9}
}

@Article{Zeiher2016NPHYS,
author={Zeiher, Johannes
and van Bijnen, Rick
and Schau{\ss}, Peter
and Hild, Sebastian
and Choi, Jae-yoon
and Pohl, Thomas
and Bloch, Immanuel
and Gross, Christian},
title={Many-body interferometry of a Rydberg-dressed spin lattice},
journal={Nature Physics},
year={2016},
month={Dec},
day={01},
volume={12},
number={12},
pages={1095-1099},
issn={1745-2481},
doi={10.1038/nphys3835},
url={https://doi.org/10.1038/nphys3835}
}

@article{Zeiher2017PRX,
  title = {Coherent Many-Body Spin Dynamics in a Long-Range Interacting Ising Chain},
  author = {Zeiher, Johannes and Choi, Jae-yoon and Rubio-Abadal, Antonio and Pohl, Thomas and van Bijnen, Rick and Bloch, Immanuel and Gross, Christian},
  journal = {Phys. Rev. X},
  volume = {7},
  issue = {4},
  pages = {041063},
  numpages = {6},
  year = {2017},
  month = {Dec},
  publisher = {American Physical Society},
  doi = {10.1103/PhysRevX.7.041063},
  url = {https://link.aps.org/doi/10.1103/PhysRevX.7.041063}
}

@Article{Qiao2025Nature,
author={Qiao, Mu
and Emperauger, Gabriel
and Chen, Cheng
and Homeier, Lukas
and Hollerith, Simon
and Bornet, Guillaume
and Martin, Romain
and G{\'e}ly, Bastien
and Klein, Lukas
and Barredo, Daniel
and Geier, Sebastian
and Chiu, Neng-Chun
and Grusdt, Fabian
and Bohrdt, Annabelle
and Lahaye, Thierry
and Browaeys, Antoine},
title={Realization of a doped quantum antiferromagnet in a Rydberg tweezer array},
journal={Nature},
year={2025},
month={Aug},
day={01},
volume={644},
number={8078},
pages={889-895},
issn={1476-4687},
doi={10.1038/s41586-025-09377-1},
url={https://doi.org/10.1038/s41586-025-09377-1}
}

@article{Geiger2018PRL,
  title = {Observation and Uses of Position-Space Bloch Oscillations in an Ultracold Gas},
  author = {Geiger, Zachary A. and Fujiwara, Kurt M. and Singh, Kevin and Senaratne, Ruwan and Rajagopal, Shankari V. and Lipatov, Mikhail and Shimasaki, Toshihiko and Driben, Rodislav and Konotop, Vladimir V. and Meier, Torsten and Weld, David M.},
  journal = {Phys. Rev. Lett.},
  volume = {120},
  issue = {21},
  pages = {213201},
  numpages = {6},
  year = {2018},
  month = {May},
  publisher = {American Physical Society},
  doi = {10.1103/PhysRevLett.120.213201},
  url = {https://link.aps.org/doi/10.1103/PhysRevLett.120.213201}
}

@article{Lu2012PRL,
  title = {Quantum Degenerate Dipolar Fermi Gas},
  author = {Lu, Mingwu and Burdick, Nathaniel Q. and Lev, Benjamin L.},
  journal = {Phys. Rev. Lett.},
  volume = {108},
  issue = {21},
  pages = {215301},
  numpages = {5},
  year = {2012},
  month = {May},
  publisher = {American Physical Society},
  doi = {10.1103/PhysRevLett.108.215301},
  url = {https://link.aps.org/doi/10.1103/PhysRevLett.108.215301}
}

@article{Aikawa2014PRL,
  title = {Reaching Fermi Degeneracy via Universal Dipolar Scattering},
  author = {Aikawa, K. and Frisch, A. and Mark, M. and Baier, S. and Grimm, R. and Ferlaino, F.},
  journal = {Phys. Rev. Lett.},
  volume = {112},
  issue = {1},
  pages = {010404},
  numpages = {5},
  year = {2014},
  month = {Jan},
  publisher = {American Physical Society},
  doi = {10.1103/PhysRevLett.112.010404},
  url = {https://link.aps.org/doi/10.1103/PhysRevLett.112.010404}
}

@article{Chomaz2023RPP,
  author={Chomaz, L. and Ferrier-Barbut, I. and Ferlaino, F. and Laburthe-Tolra, B. and Lev, B. L. and Pfau, T.},
  title={Dipolar physics: a review of experiments with magnetic quantum gases},
  journal={Rep. Prog. Phys.}, volume={86}, pages={026401}, year={2023},
  doi={10.1088/1361-6633/aca814}
}

@Article{Barral2024NC,
author={Barral, Pierre
and Cantara, Michael
and Du, Li
and Lunden, William
and de Hond, Julius
and Jamison, Alan O.
and Ketterle, Wolfgang},
title={Suppressing dipolar relaxation in thin layers of dysprosium atoms},
journal={Nature Communications},
year={2024},
month={Apr},
day={26},
volume={15},
number={1},
pages={3566},
issn={2041-1723},
doi={10.1038/s41467-024-47260-1},
}

@article{Maier2015PRA,
  title = {Broad universal Feshbach resonances in the chaotic spectrum of dysprosium atoms},
  author = {Maier, T. and Ferrier-Barbut, I. and Kadau, H. and Schmitt, M. and Wenzel, M. and Wink, C. and Pfau, T. and Jachymski, K. and Julienne, P. S.},
  journal = {Phys. Rev. A},
  volume = {92},
  issue = {6},
  pages = {060702(R)},
  numpages = {5},
  year = {2015},
  month = {Dec},
  publisher = {American Physical Society},
  doi = {10.1103/PhysRevA.92.060702},
  url = {https://link.aps.org/doi/10.1103/PhysRevA.92.060702}
}

@Article{Frisch2014Nature,
author={Frisch, Albert
and Mark, Michael
and Aikawa, Kiyotaka
and Ferlaino, Francesca
and Bohn, John L.
and Makrides, Constantinos
and Petrov, Alexander
and Kotochigova, Svetlana},
title={Quantum chaos in ultracold collisions of gas-phase erbium atoms},
journal={Nature},
year={2014},
month={Mar},
day={01},
volume={507},
number={7493},
pages={475-479},
issn={1476-4687},
doi={10.1038/nature13137},
url={https://doi.org/10.1038/nature13137}
}

@article{FerrierBarbut2016PRL,
  title = {Observation of Quantum Droplets in a Strongly Dipolar Bose Gas},
  author = {Ferrier-Barbut, Igor and Kadau, Holger and Schmitt, Matthias and Wenzel, Matthias and Pfau, Tilman},
  journal = {Phys. Rev. Lett.},
  volume = {116},
  issue = {21},
  pages = {215301},
  numpages = {6},
  year = {2016},
  month = {May},
  publisher = {American Physical Society},
  doi = {10.1103/PhysRevLett.116.215301},
  url = {https://link.aps.org/doi/10.1103/PhysRevLett.116.215301}
}

@article{Anderegg2021Science,
author = {Loïc Anderegg  and Sean Burchesky  and Yicheng Bao  and Scarlett S. Yu  and Tijs Karman  and Eunmi Chae  and Kang-Kuen Ni  and Wolfgang Ketterle  and John M. Doyle },
title = {Observation of microwave shielding of ultracold molecules},
journal = {Science},
volume = {373},
number = {6556},
pages = {779-782},
year = {2021},
doi = {10.1126/science.abg9502}}

@Article{Schindewolf2022Nature,
author={Schindewolf, Andreas
and Bause, Roman
and Chen, Xing-Yan
and Duda, Marcel
and Karman, Tijs
and Bloch, Immanuel
and Luo, Xin-Yu},
title={Evaporation of microwave-shielded polar molecules to quantum degeneracy},
journal={Nature},
year={2022},
month={Jul},
day={01},
volume={607},
number={7920},
pages={677-681},
issn={1476-4687},
doi={10.1038/s41586-022-04900-0},
url={https://doi.org/10.1038/s41586-022-04900-0}
}

@Article{Yan2013Nature,
author={Yan, Bo
and Moses, Steven A.
and Gadway, Bryce
and Covey, Jacob P.
and Hazzard, Kaden R. A.
and Rey, Ana Maria
and Jin, Deborah S.
and Ye, Jun},
title={Observation of dipolar spin-exchange interactions with lattice-confined polar molecules},
journal={Nature},
year={2013},
month={Sep},
day={01},
volume={501},
number={7468},
pages={521-525},
issn={1476-4687},
doi={10.1038/nature12483},
url={https://doi.org/10.1038/nature12483}
}

@article{dePaz2013PRL,
  title = {Nonequilibrium Quantum Magnetism in a Dipolar Lattice Gas},
  author = {de Paz, A. and Sharma, A. and Chotia, A. and Mar\'echal, E. and Huckans, J. H. and Pedri, P. and Santos, L. and Gorceix, O. and Vernac, L. and Laburthe-Tolra, B.},
  journal = {Phys. Rev. Lett.},
  volume = {111},
  issue = {18},
  pages = {185305},
  numpages = {5},
  year = {2013},
  month = {Oct},
  publisher = {American Physical Society},
  doi = {10.1103/PhysRevLett.111.185305},
  url = {https://link.aps.org/doi/10.1103/PhysRevLett.111.185305}
}

@article{Patscheider2020PRR,
  title = {Controlling dipolar exchange interactions in a dense three-dimensional array of large-spin fermions},
  author = {Patscheider, A. and Zhu, B. and Chomaz, L. and Petter, D. and Baier, S. and Rey, A.-M. and Ferlaino, F. and Mark, M. J.},
  journal = {Phys. Rev. Res.},
  volume = {2},
  issue = {2},
  pages = {023050},
  numpages = {16},
  year = {2020},
  month = {Apr},
  publisher = {American Physical Society},
  doi = {10.1103/PhysRevResearch.2.023050},
  url = {https://link.aps.org/doi/10.1103/PhysRevResearch.2.023050}
}

@article{Ferrari2006PRL,
  title = {Long-Lived Bloch Oscillations with Bosonic Sr Atoms and Application to Gravity Measurement at the Micrometer Scale},
  author = {Ferrari, G. and Poli, N. and Sorrentino, F. and Tino, G. M.},
  journal = {Phys. Rev. Lett.},
  volume = {97},
  issue = {6},
  pages = {060402},
  numpages = {4},
  year = {2006},
  month = {Aug},
  publisher = {American Physical Society},
  doi = {10.1103/PhysRevLett.97.060402},
  url = {https://link.aps.org/doi/10.1103/PhysRevLett.97.060402}
}

@article{Trotzky2008Science,
author = {S. Trotzky  and P. Cheinet  and S. Fölling  and M. Feld  and U. Schnorrberger  and A. M. Rey  and A. Polkovnikov  and E. A. Demler  and M. D. Lukin  and I. Bloch },
title = {Time-Resolved Observation and Control of Superexchange Interactions with Ultracold Atoms in Optical Lattices},
journal = {Science},
volume = {319},
number = {5861},
pages = {295-299},
year = {2008},
doi = {10.1126/science.1150841},
URL = {https://www.science.org/doi/abs/10.1126/science.1150841}}

@article{Boll2016Science,
author = {Martin Boll  and Timon A. Hilker  and Guillaume Salomon  and Ahmed Omran  and Jacopo Nespolo  and Lode Pollet  and Immanuel Bloch  and Christian Gross },
title = {Spin- and density-resolved microscopy of antiferromagnetic correlations in Fermi-Hubbard chains},
journal = {Science},
volume = {353},
number = {6305},
pages = {1257-1260},
year = {2016},
doi = {10.1126/science.aag1635},
URL = {https://www.science.org/doi/abs/10.1126/science.aag1635}}

@Article{Mazurenko2017Nature,
author={Mazurenko, Anton
and Chiu, Christie S.
and Ji, Geoffrey
and Parsons, Maxwell F.
and Kan{\'a}sz-Nagy, M{\'a}rton
and Schmidt, Richard
and Grusdt, Fabian
and Demler, Eugene
and Greif, Daniel
and Greiner, Markus},
title={A cold-atom Fermi--Hubbard antiferromagnet},
journal={Nature},
year={2017},
month={May},
day={01},
volume={545},
number={7655},
pages={462-466},
issn={1476-4687},
doi={10.1038/nature22362},
url={https://doi.org/10.1038/nature22362}
}

@Article{Bakr2009Nature,
author={Bakr, Waseem S.
and Gillen, Jonathon I.
and Peng, Amy
and F{\"o}lling, Simon
and Greiner, Markus},
title={A quantum gas microscope for detecting single atoms in a Hubbard-regime optical lattice},
journal={Nature},
year={2009},
month={Nov},
day={01},
volume={462},
number={7269},
pages={74-77},
issn={1476-4687},
doi={10.1038/nature08482},
url={https://doi.org/10.1038/nature08482}
}

@Article{Sherson2010Nature,
author={Sherson, Jacob F.
and Weitenberg, Christof
and Endres, Manuel
and Cheneau, Marc
and Bloch, Immanuel
and Kuhr, Stefan},
title={Single-atom-resolved fluorescence imaging of an atomic Mott insulator},
journal={Nature},
year={2010},
month={Sep},
day={01},
volume={467},
number={7311},
pages={68-72},
issn={1476-4687},
doi={10.1038/nature09378},
url={https://doi.org/10.1038/nature09378}
}

@article{Baier2016Science,
  author    = {Baier, S. and Mark, M. J. and Petter, D. and Aikawa, K. and Chomaz, L. and Cai, Z. and Baranov, M. and Zoller, P. and Ferlaino, F.},
  title     = {Extended {Bose}-{Hubbard} Models with Ultracold Magnetic Atoms},
  journal   = {Science},
  volume    = {352},
  number    = {6282},
  pages     = {201--205},
  year      = {2016},
  publisher = {American Association for the Advancement of Science},
  doi       = {10.1126/science.aac9812}
}

\end{document}